\documentclass[useAMS,usenatbib]{mn2e}
\usepackage{hyperref}
\usepackage{epsfig}

\def\gsim{~\rlap{$>$}{\lower 1.0ex\hbox{$\sim$}}}
\def\lsim{~\rlap{$<$}{\lower 1.0ex\hbox{$\sim$}}}
\def\G{{\rm G}}

\def\d{{\rm d}}

\newcommand{\pcite}[1]{(\citealt{#1})}

\title{The Origin of the Hubble Sequence in ${\Lambda}$CDM Cosmology}
\author[Andrew J. Benson and Nick Devereux]{Andrew J. Benson$^1$ and Nick Devereux$^2$\\
$^1$Mail Code 130-33, California Institute of Technology, Pasadena, CA~91125, U.S.A. (e-mail: {\tt abenson@caltech.edu})\\
$^2$Department of Physics, Embry-Riddle Aeronautical University, Prescott, AZ~86301-3720, U.S.A. (e-mail: {\tt devereux@erau.edu})}

\begin{document}

\maketitle

\begin{abstract}

The {\sc Galform} semi-analytic model of galaxy formation is used to explore the mechanisms 
primarily responsible for the three types of galaxies
seen in the local universe:  bulge, bulge+disk and disk, identified with the visual morphological types E, S0/a-Sbc, and Sc-Scd, respectively. With a suitable choice of parameters the {\sc Galform} model can accurately reproduce the observed local $K_{s}$-band luminosity function (LF) for galaxies split by visual morphological type. The successful set of model parameters
is used to populate the Millennium Simulation with 9.4 million galaxies
and their dark matter halos. The resulting catalogue is then used to explore the evolution of 
galaxies through cosmic history. The model predictions concur with recent observational results including the galaxy merger rate, the star formation rate and the seemingly anti-hierarchical evolution of ellipticals. However, the model also predicts significant
evolution of the elliptical galaxy LF that is not observed. The discrepancy raises the possibility that samples of $z~{\sim}$ 1 galaxies which have been selected using colour and morphological criteria
may be contaminated with galaxies that are not actually ellipticals.
\end{abstract}


\section{Introduction}

The current paradigm, constrained in part by the $K_{s}$-band (hereafter $K$-band)  luminosity function (LF) (e.g., \citealt{benson_what_2003}), has galaxy disks forming through a combination of cold gas accretion \citep{weinberg_galaxy_2004} and feedback \citep{oppenheimer_cosmological_2006-1}, with bulges resulting from 
mergers \citep[e.g.,][]{barnes_dynamics_1992, hopkins_cosmological_2008, masjedi_growth_2008}. Thus, the diversity of galaxy morphologies seen today represents the culmination of multiple evolutionary paths.  These end points in galaxy evolution are captured in a taxonomy devised by  \citet{hubble_realm_1936} and refined by \citet{de_vaucouleurs_classification_1959} and \cite{de_vaucouleurs_third_1991} that is based on the relative prominence of the stellar bulge and the degree of resolution of the spiral arms.  Despite attempts to emulate this classification scheme using quantitative proxies such as colours \citep[e.g.,][]{strateva_color_2001}, concentration-asymmetry indices \citep[e.g.,][]{watanabe_digital_1985, abraham_galaxy_1996, bell_optical_2003}, the Gini-M20 classifiers \citep{lotz_new_2004} and parametric methods involving bulge/disk decompositions \citep{schade_evolution_1996, ratnatunga_disk_1999, simard_deep_2002} there are no quantitative measures that uniquely segregate galaxies into the morphological groups represented by the de Vaucouleurs numerical {\tt T}  system \citep[e.g.,][]{graham_inclination-_2008}. Thus, proxies are merely a useful stopgap until the difficult and time consuming task of assigning visible morphologies to the ever-growing number of cataloged galaxies can be accomplished\footnote{See the Galaxy Zoo Project, \href{http://www.galaxyzoo.org/}{\tt http://www.galaxyzoo.org/}}.

Fortunately, the time-consuming task of assigning visible morphologies has 
now been completed for the vast majority of nearby galaxies.  The principal aim of this paper is therefore to use the $K$-band LF for nearby galaxies \citep{devereux_nearby_2009} to constrain models of their formation and evolution. The $K$-band LF  is of particular interest because of its relevance to understanding galaxy evolution in the context of Lambda Cold Dark Matter (${\Lambda}$CDM) cosmology; at $z=0$, the $K$-band LF traces the stellar mass accumulated in galaxies at a wavelength where interstellar extinction is minimal \citep{devereux_infrared_1987, bell_stellar_2000, bell_optical_2003}. Additionally, the functional forms for the $K$-band LFs distinguish between {\it bulge} dominated and  {\it disk}  dominated systems which suggests that at least two quite distinct galaxy formation mechanisms are at work to produce the diversity of morphological types seen in the local universe. Previous attempts to model the LFs in the context of hierarchical clustering scenarios \citep[e.g.,][]{cole_hierarchical_2000,benson_what_2003} have focussed on the {\it Universal} or {\it Total} $K$-band LF, which is the sum over all galaxy types, because they are based on samples for which there are no pre-existing morphological assignments. In this paper, the $K$-band LFs are modeled for galaxies segregated by visible morphology in order to better understand:
1) the origin and evolution of bulgeless disk galaxies, represented in the observable universe by Sc-Scd galaxies; 2) the role of secular processes in building the bulges of spiral (S0/a---Sbc) and lenticular galaxies and; 3) the galaxy merger history that led to the formation of elliptical galaxies.
Our analysis differs from \cite{parry_galaxy_2009} in that the GALFORM model predictions
are constrained by the observed $K$-band LFs for galaxies of different morphological type. 
Those model predictions are then compared with recent observational results including
the type averaged galaxy merger rate and the type specific star formation rate, both 
quantified as a function of redshift.

Observationally, there is a weak dependence of $K$-band B/T on visual morphology \pcite{graham_inclination-_2008} which is used
in this paper to segregate the model galaxies into three broad groups
representing the three types of galaxies
seen in the local universe, namely bulge, bulge+disk and disk.\footnote{In principle, other outputs from the model could be used to provide additional information on morphological type. However \protect\cite{graham_inclination-_2008} show that bulge/disk flux ratio is the most useful discriminant as it varies more, as a function of morphological type, than other indicators, such as bulge/disk size ratio, for example.}
These three groups are identified with the visual morphological types E, S0/a-Sbc, and Sc-Scd, respectively. Following \cite{graham_inclination-_2008}, these three broad groups are identified in the models by adopting the range of B/T ratio listed in Table~\ref{tb:BTMap}. The table also indicates the distribution of B/T within each morphological class, illustrating that most disk galaxies (including S0's) have a rather low B/T ratio. For example, 80\% of S0 and later type galaxies brighter than $M_{\rm K}-5\log_{10}{\it h} = -21$ have B/T $\le$ 0.54. If the magnitude limit is lowered to $M_{\rm K}-5\log_{10}{\it h} = -19$, we find that 80\% of such galaxies have B/T  $\le$ 0.35. This is in agreement with the broad conclusion of \cite{graham_inclination-_2008} that most disk galaxies (including S0's) have low B/T ratios (e.g. B/T $<$1/3).

\begin{table}
 \label{tb:BTMap}
 \begin{center}
 \begin{tabular}{lcccc}
 \hline
 {\bf Morphological} & & \multicolumn{3}{c}{{\bf B/T distribution}} \\
 {\bf class} & {\bf B/T Range} & 20\% & 50\% & 80\% \\
 \hline
 E      & $0.92 <   B/T_{\rm K} \le 1.00$ & 0.99 & 1.00 & 1.00 \\
 S0-Sbc & $0.11 <   B/T_{\rm K} \le 0.92$ & 0.22 & 0.39 & 0.65 \\
 Sc-Scd & $0.00 \le B/T_{\rm K} \le 0.11$ & 0.00 & 0.01 & 0.05 \\
 \hline
 \end{tabular}
 \end{center}
 \caption{The Table presents the mapping between morphological type and K-band bulge-to-total ratio required for our best-fit model to reproduce the relative abundances of each class in the interval $-23.5 < M_{\rm K}-5\log_{10}h \le -23$. Additionally, the final three columns show the 
percentage of galaxies in each class with $M_{\rm K}-5\log h < -21$
and  B/T ratios less than the value indicated. For example, 80\% of bright S0-Sbc galaxies have a B/T less than 0.65.}
\end{table}

\section{Model}\label{sec:Model}

The {\sc Galform} semi-analytic model has been described most recently by  \cite{bower_breakinghierarchy_2006} and the reader is referred to that paper for a full description. In the model, galaxies consist of an exponential disk containing both gas and stars and a spheroidal stellar component---the morphology of the model 
galaxies is therefore determined by the relative amount of light in each component. A few key points regarding morphological evolution of model galaxies bear reiterating however:
\begin{itemize}
 \item Galaxies initially form through the cooling of gas from their surrounding dark matter halo. This gas is assumed to conserve its angular momentum during collapse and to therefore form a disk. As such, a galaxy formed in isolation will always be a pure disk galaxy unless it is sufficiently self-gravitating to become unstable.
 \item Disks which become too self-gravitating become unstable to the formation of global perturbations and therefore destroy themselves. Instability is judged using the criterion proposed by \cite{efstathiou_stability_1982}. For a disk to be stable it must satisfy
\begin{equation}
{v_{\rm D} \over (\G M_{\rm D}/R_{\rm D})^{1/2}} > \epsilon,
\end{equation}
where $v_{\rm }$ is the circular velocity of the disk at its half-mass radius, $R_{\rm D}$ and $M_{\rm D}$ is the mass of the disk. The value of $\epsilon$ for purely stellar disks was found to be $1.1$ by \cite{efstathiou_stability_1982}, although lower values may be appropriate for gas rich disks \pcite{christodoulou_new_1995}. We treat $\epsilon$ as a free parameter. If a disk is deemed to be unstable according to this criterion, the disk is destroyed, with any stars and gas present forming a spheroidal system in which the gas is then rapidly turned into stars\footnote{Since the treatment of this instability process in semi-analytic models is somewhat uncertain, an alternative and less dramatic implementation is presented in the Appendix~\protect\ref{app:altDiskInstab}. The
results do not change the qualitative conclusions of this paper.}. The end result is a purely spheroidal (a.k.a. elliptical) system.
 \item Mergers between galaxies can also create a spheroidal component. Galaxy-galaxy mergers are driven by dynamical friction on a satellite galaxy (and its halo) orbiting in a larger dark matter halo. Merging timescales are computed using a modified form of Chandrasekhar's dynamical friction equation \pcite{lacey_merger_1993}. The timescales 
 are scaled by $\tau^0_{\rm mrg}$, a parameter of the model, to allow some adjustment of merger rates. Mergers are split into two categories:
 \begin{itemize}
  \item \emph{Minor Mergers}: These are mergers with $M_2/M_1<f_{\rm ellip}$ where $M_1$ and $M_2<M_1$ are the baryonic masses of the galaxies involved in the merger and $f_{\rm ellip}$ is a parameter of the model (see below). In a minor merger, the stars from the secondary (lower mass) galaxy are added to the spheroidal component of the primary (more massive) galaxy while any gas from the secondary is added to the disk of the primary. A galaxy can therefore grow a spheroidal component through the cumulative effects of many minor mergers.
  \item \emph{Major Mergers}: These are mergers with $M_2/M_1\ge f_{\rm ellip}$ and cause the destruction of any prexisting stellar disks, such that all stars from the merging galaxies become part of the spheroidal component in the merger remnant. The fate of any gas in the merging galaxies also depends on the merger mass ratio:
   \item If $M_2/M_1<f_{\rm burst}$, where $f_{\rm burst}$ is a parameter of the model, then the merger does not trigger a burst of star formation and any gas forms a disk in the merger remnant (and may subsequently form stars in that disk).
   \item If $M_2/M_1\ge f_{\rm burst}$ then the merger triggers a burst of star formation and any gas present in the merging galaxies is turned into stars on a short timescale and becomes part of the spheroidal component of the merger remnant.
  \end{itemize}
  \end{itemize}
 \begin{itemize}
 \item Once a spheroidal component has formed, subsequent gas cooling can cause a disk to regrow around the spheroid, resulting in a galaxy with intermediate bulge-to-total ratio.
\item The model includes supernovae feedback which is assumed to drive gas out of a galaxy at a rate proportional to the star formation rate: $\dot{M}_{\rm out} = \beta \dot{M}_\star$, where $\beta=(V_{\rm hot}/V)^{\alpha_{\rm hot}}$, $V$ is the circular speed of the galaxy and $V_{\rm hot}$ and $\alpha_{\rm hot}$ are parameters of the model. $V_{\rm hot}$ may differ for quiescent star formation in disks and bursts of star formation.
\end{itemize}

\begin{figure}
 \begin{center}
 \includegraphics[width=80mm,viewport=0mm 55mm 205mm 245mm,clip]{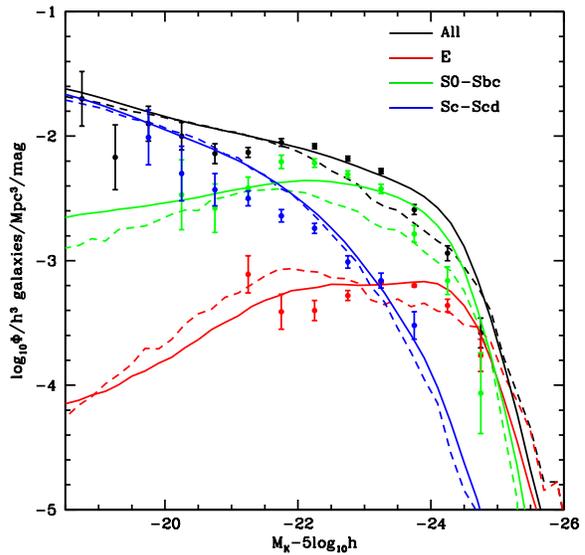}
 \end{center}
 \caption{Morphologically selected K-band LFs for the best fit model (Table 2) are shown by solid lines and are compared to the observed LFs of \protect\cite{devereux_nearby_2009} shown by the points. Colours indicate morphological class as indicated by the key in the figure. Galaxies are split into three morphological classes according to Table 1. For reference, dashed lines show results from the \protect\cite{bower_breakinghierarchy_2006} model (i.e. without the parameter modifications introduced in this work) using the same cuts on B/T to define morphological classes.}
 \label{fig:BestLF}
\end{figure}

The \cite{bower_breakinghierarchy_2006} model was not constructed using morphological data as a constraint. 
Our initial goal, therefore, is to adjust the parameters of the \cite{bower_breakinghierarchy_2006} model so that it reproduces the observed morphologically segregated LFs of \cite{devereux_nearby_2009}, shown in Figure~\ref{fig:BestLF}. A search for an alternative set of 
values for the six model parameters that match the observed morphologically selected LFs yielded the values listed in Table~\ref{tb:Params}. The best fit model LFs (lines) are compared with the observed LFs from \cite{devereux_nearby_2009} (points) in Figure~\ref{fig:BestLF}. Morphological class is indicated by colour in the figure. The relative abundance of each type in the model is forced to agree with the data in the interval $-23.5 < M_{\rm K}-5\log_{10}h \le -23.0$ (this is how the mapping between B/T and morphological class listed in Table~\ref{tb:BTMap} was determined) but the morphological mix is not enforced outside this magnitude range. Also shown, for reference, are the results from the unmodified \cite{bower_breakinghierarchy_2006} model (dashed lines). The comparison 
shows that the model results are quite similar.

The model LFs for ellipticals, lenticulars and bulge-dominated spirals are peaked and decline toward both higher and lower luminosities. In contrast, the Sc-Scd class continues to 
rise towards low luminosities and dominates the total LF for 
$M_{\rm K} - 5 \log_{10} h > -22$. 
The black line and points indicate the total (i.e. all morphological types) LF. Not surprisingly, the model prediction here is also in good agreement with the data. This is to be expected
however,  since the model parameters are 
optimised to yield a satisfactory representation of the
observed LFs for each of the three broad morphological types and the total LF is just the sum of these. Additionally, \cite{bower_breakinghierarchy_2006} have already demonstrated that the model can fit the observed total K-band LF. The parameters yielding the best-fit model (Table~\ref{tb:Params}) are similar to those adopted by  \cite{bower_breakinghierarchy_2006}: disk instability estimates are unchanged, but the rate of galaxy merging has been reduced by a factor of 2, a lower mass ratio in mergers is required to disrupt stellar disks, a higher ratio is required to trigger a burst, while the feedback in bursts, but not disks, has been weakened.

\begin{table}
 \caption{Parameters of the best-fit model.}
 \label{tb:Params}
 \begin{center}
 \begin{tabular}{lcc}
 \hline
  & \multicolumn{2}{c}{\bf Value}\\
  \cline{2-3} 
 {\bf Parameter} & {\bf Best Fit} & {\bf Bower et al. 2006} \\
 \hline
 $\epsilon$ & 0.8 & 0.8 \\
 $\tau_{\rm 0 mrg}$ & 3.0 & 1.5 \\
 $f_{\rm ellip}$ & 0.15 & 0.3 \\
 $f_{\rm burst}$ & 0.2 & 0.1 \\
 $V_{\rm hot,disk}$ & 485km/s & 485km/s \\
 $V_{\rm hot,burst}$ & 300km/s & 485km/s \\
 \hline
 \end{tabular}
 \end{center}
\end{table}

\section{Results}\label{sec:Results}

Having found a successful set of model parameters, they
are used to populate the Millennium Simulation \pcite{springel_simulations_2005} with galaxies. To do this, dark matter halo merger trees are extracted from the Millennium Simulation using the techniques described by \cite{harker_marked_2006}. The trees are then populated with galaxies in the manner described by \cite{bower_breakinghierarchy_2006}, but using the model parameters found in the previous section. This results in a catalogue of 9.4 million galaxies
and their dark matter halos. The catalogue provides the magnitudes, morphologies and other physical characteristics for all of the galaxies, including their evolution through cosmic history. The catalogue can therefore be used to explore galaxy evolution. Our goal is to reveal the key physical processes at
work that produce the three broad types of galaxies in the observable universe, namely bulges, bulges+disks and disks. The resolution of the Millennium Simulation, as in any N-body simulation, is limited. In Appendix~\ref{app:ResStudy} a study of the effects of resolution demonstrates that the morphological features of the model galaxies are well converged at the resolution of the Millennium Simulation.

\subsection{Trends with Morphological Class}

\subsubsection{Dark Matter Halo Properties}

\begin{figure}
 \includegraphics[width=80mm,viewport=0mm 55mm 205mm 245mm,clip]{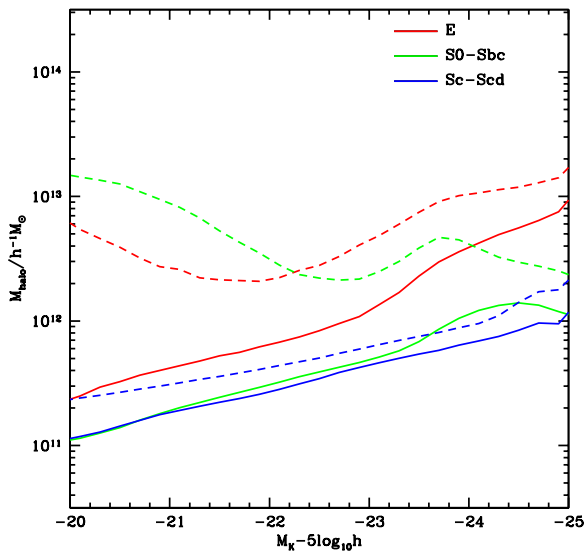}
 \caption{The median halo mass in which galaxies formed (solid lines) and the mass of their current halo at $z=0$ (dashed lines) as a function of K-band luminosity and morphological class (colour coding as in Fig.~\protect\ref{fig:BestLF}). The current halo mass is the same as the mass of the halo in which the galaxy formed for central galaxies but may be significantly larger for satellite galaxies.}
 \label{fig:HaloMassBest}
\end{figure}

The relationship between galaxy luminosity and dark matter halo mass is illustrated in Fig.~\ref{fig:HaloMassBest} as a function of morphological type. The solid lines in Fig.~\ref{fig:HaloMassBest} indicate the mass of the halo in which a galaxy formed, defined here following \cite{cole_hierarchical_2000} such that halos are labelled as ``reforming'' each time they undergo a doubling of their mass.  As may be expected there is a correlation between galaxy luminosity and the median mass of the dark matter 
halo in which such galaxies formed. The slope of the relation is quite flat however, showing a range of a factor of ${\sim}$20 in halo mass over a factor of ${\sim}$100 range in luminosity. 

There is little dependence of median formation halo mass on morphological type, except for the E class which form in halos around a factor 
of ${\sim}$ 3 more massive for a given luminosity. This may be expected for the E class if such galaxies are built up via merging rather than continuous gas accretion. To attain the same luminosity as its accretion-fueled counterpart, a merger-built galaxy must form in a more massive halo so that the rate of merging will be enhanced.

The dashed lines in Fig.~\ref{fig:HaloMassBest} indicate the mass of the 
dark matter halo in which a galaxy lives at $z = 0$. In the case of satellite galaxies, this is the mass of the isolated dark matter halo in which they orbit, \emph{not} the mass of the subhalo within which they reside. 
The difference between the dashed and solid lines 
is therefore indicative of how much mass has accumulated in the dark matter halo since galaxies formed. Dark matter halos may
grow through mergers of satellites and by accretion. With the exception of Sc-Scd galaxies, the mass of the halo in which galaxies live at $z = 0$ greatly exceeds the mass of halo in which they formed, particularly for the lower luminosity systems (a simple consequence of the fact that the low mass halos in which these faint galaxies form are highly likely to be subsumed into much more massive halos by $z=0$ by becoming satellites of those halos).

\begin{figure}
 \includegraphics[width=80mm,viewport=0mm 55mm 205mm 245mm,clip]{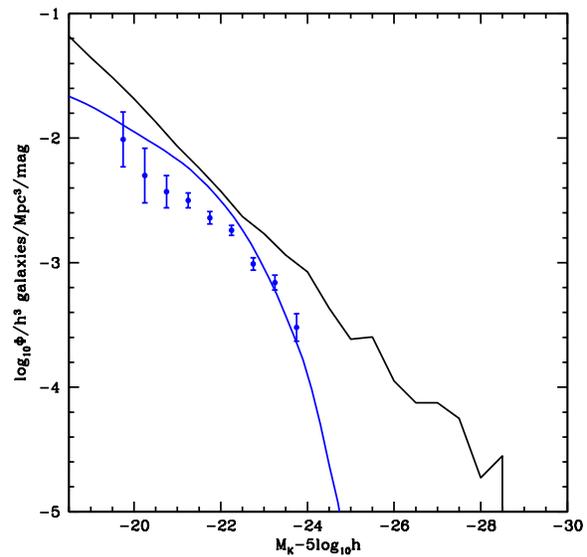}
 \caption{The Sc-Scd K-band LF (blue line and points) versus the median formation halo mass function
 (black line) scaled by a M/L ratio. }
 \label{fig:LFLateBest}
\end{figure}

Interestingly, Sc-Scd galaxies of all luminosities occupy
dark matter halos that are only about a factor of two more massive than those in which they formed. Thus, disk dominated galaxies must have avoided major mergers by virtue of living in dark matter halos having rather quiet merging histories. There are a wide variety of possible merger histories for halos of a given mass \pcite{stewart_merger_2008} including some fraction which will have avoided major mergers. It is known that halos in low density environments experience lower merger rates than their counterparts in high density environments \pcite{fakhouri_environmental_2009} and so are more likely to have a quiet merger history. Thus, Sc-Scd type galaxies are distinguished from the other morphological types having formed preferentially in low density environments. 65\% of Sc-Scd galaxies brighter than $M_{\rm K} -5 \log_{10} h = -20$ are central galaxies, ie. they are the most massive galaxy in their halo,
and they grow primarily through gas accretion and not mergers. Consequently, the Sc-Scd K-band 
luminosity function has a similar slope to the median formation halo mass function as illustrated in
Fig.~\ref{fig:LFLateBest}

\subsubsection{Timescales}

While interesting from a theoretical standpoint, dark matter halo masses cannot be easily measured for most galaxies. (Although weak lensing \pcite{mandelbaum_density_2006} and clustering \pcite{mandelbaum_halo_2009} measures can constrain halo mass for classes of galaxies.) Greater utility is provided by examining observable quantities more directly related to the galaxies themselves, beginning with timescales.

\begin{figure}
 \includegraphics[width=80mm,viewport=0mm 55mm 205mm 245mm,clip]{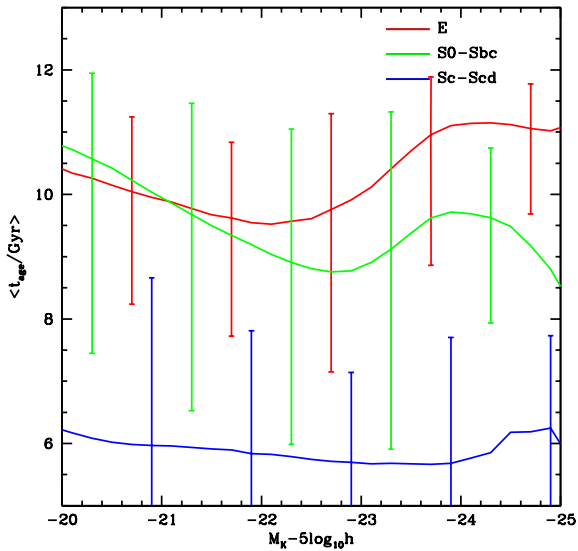}
 \caption{The stellar mass-weighted age of galaxies as a function of K-band magnitude and morphological class (colour coding as in Fig.~\protect\ref{fig:BestLF}). Solid lines indicate the median ages for galaxies in each morphological class while error bars indicate the $\pm 1\sigma$ interval of the distribution of ages.}
 \label{fig:AgesBest}
\end{figure}

Figure~\ref{fig:AgesBest} shows the median stellar mass-weighted age of galaxies, defined as
\begin{equation}
 \langle t_{\rm age} \rangle = {\int_0^{t_0} (t_0-t)\dot{\rho}_\star(t) \d t \over \int_0^{t_0} \dot{\rho}_\star(t) \d t }
\end{equation}
where $t_0$ is the current age of the Universe and $\dot{\rho}_\star(t)$ is the star formation rate in a galaxy at time $t$, split by morphological type as a function of their $K$-band magnitude. Disk dominated galaxies (blue line) are distinguished once again, this time by hosting a uniformly much younger stellar population than any other morphological type. This suggests that disk galaxies of all luminosities have been continuously forming stars since their formation which the model reveals is at a rate that is proportional to the baryonic mass of the galaxy. In contrast the ellipticals, lenticulars and bulge dominated spirals contain much older stellar populations with median ages of 10Gyr. This does not mean that the galaxies formed 10Gyr ago, merely that the stars now dominating the baryonic mass of the galaxies typically formed at that time. As we shall see in \S\ref{sec:zDists}, massive ellipticals were actually assembled much more recently.

\subsubsection{Bursts and Bulge Formation}

\begin{figure}
 \includegraphics[width=80mm,viewport=0mm 55mm 205mm 245mm,clip]{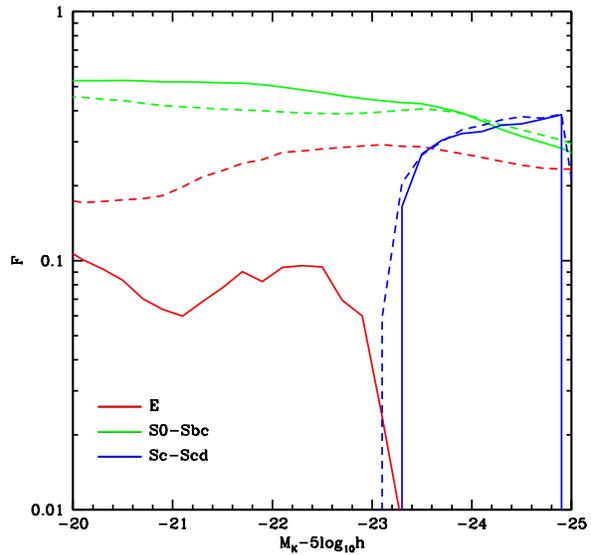}
 \caption{The fraction of a galaxy's bulge mass which was made in the most recent burst of star formation 
(solid lines) and disk instabilities (dashed lines) as a function of K-band magnitude for each morphological class (colour coding as in Fig.~\protect\ref{fig:BestLF}).}
\label{fig:BulgeMassBest}
\end{figure}

Figure~\ref{fig:BulgeMassBest} examines the fraction of a galaxy's bulge mass that
was made in the most recent burst of star formation (solid
lines) and the fraction formed through disk instabilities (the
remainder being made through minor or major mergers).
Interestingly, $\sim$30--50\% of the bulge mass in lenticulars and bulge
dominated spirals (S0/a - Sbc) resulted from disk instabilities with the
remainder created in the most recent burst of star formation, which
occurrred ${\le}$10Gyr ago, that itself could have been triggered by a disk
instability or a merger. Unfortunately, there is no way to distinguish between the two possibilties in this figure. The importance of these processes is greatly diminished in
elliptical galaxies, because they do not possess disks\footnote{The dashed lines reflect the mass 
of stars 
that were formed through disk instabilities in {\it all} progenitors. Thus the 
dashed red-line indicates that disk instabilties may have been an important mechanism for the progenitors of todays ellipticals. We note that the fraction of an elliptical's mass which formed as a result of disk instability events is sensitive to the details of the treatment of such instabilities in our model. With our standard treatment, on averge 29\% of the spheroid mass in bright ($M_{\rm K}-5\log_{10}h<-23$) ellipticals is made via disk instability events. Using a more moderate treatment of instability events in which a minimal amount of disk is converted into spheroid (see Appendix~\ref{app:altDiskInstab}) this number is reduced to 9\%. These fractions are relatively insensitive to the resolution of the merger trees. The resolution study described in Appendix~\ref{app:ResStudy} indicates that they change by only 3\% for the brightest ellipticals as the resolution of the merger trees is increased by a factor of 16. For the faintest ellipticals shown (which are closer to the resolution limit of our calculations) the fraction changes by only 7\% with the same increase in resolution.}
and in disk
dominated Sc-Scd galaxies, because they have small or non existent bulges\footnote{However, the models do indicate that disk instabilities may be an important bulge 
formation mechanism in the luminous, M$_{\rm K}-5\log_{10}h<-23$ mag, Sc-Scd galaxies.}

\begin{figure*}
 \includegraphics[width=160mm,viewport=5mm 20mm 190mm 155mm,clip]{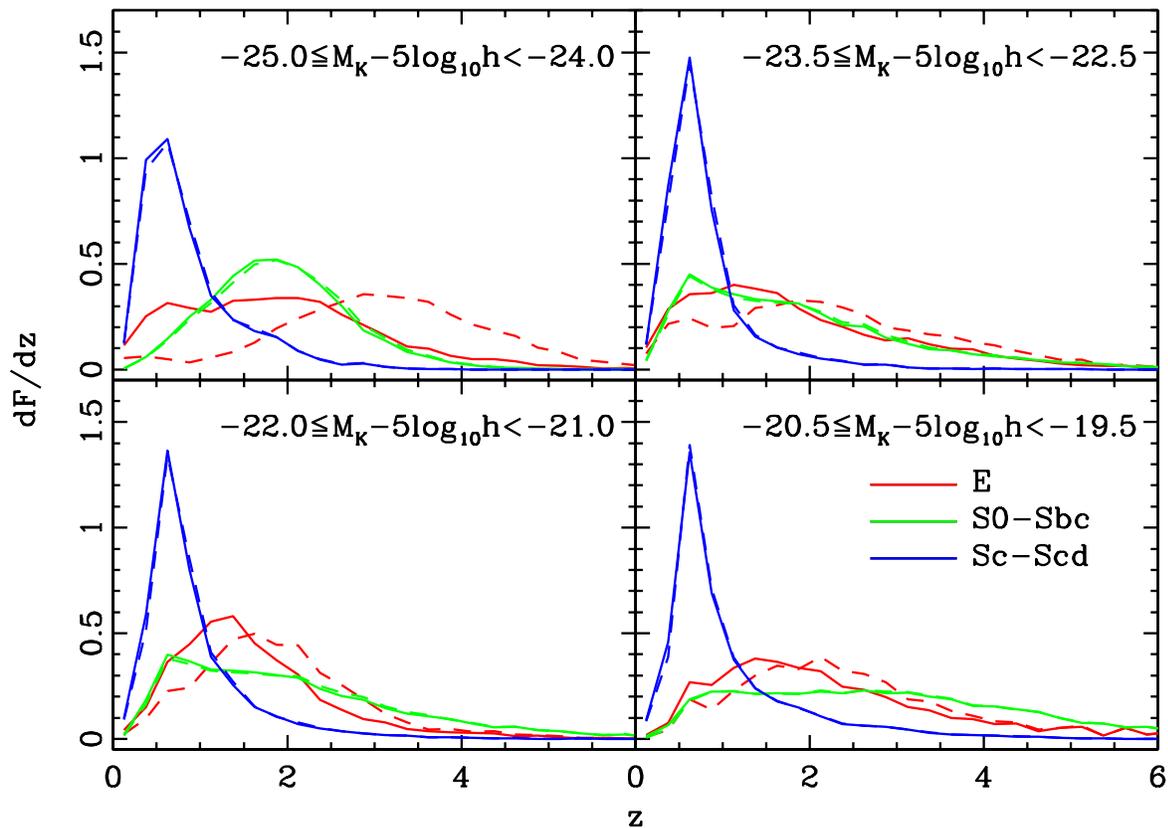}
 \caption{The normalized distributions of $z_\star$ (the redshift at which half of a galaxy's final number of stars were formed; dashed lines) and $z_{\rm assem}$ (the redshift at which half of the galaxy's final stellar mass was assembled into a single progenitor; solid lines) are shown split by morphological class (colour coding as in Fig.~\protect\ref{fig:BestLF}). Each panel corresponds to a different K-band magnitude interval as indicated in the panel. The distribution if normalized such that the integral under the curve is unity.}
 \label{fig:StarsAssem}
\end{figure*}

\subsubsection{Formation and Assembly Redshift Distributions}\label{sec:zDists}

Figure~\ref{fig:StarsAssem} shows the distribution of star formation redshifts,
$z_\star$, and assembly redshifts, $z_{\rm assem}$. The star formation redshift for a galaxy is defined to be the redshift at which half
of the final stellar mass of the galaxy has been formed. The
assembly redshift is the redshift at which half of the final
stellar mass has been assembled into a single progenitor.
Clearly for a given galaxy $z_{\rm assem} < z_\star$.
The most striking feature of these distributions is
that the most luminous -25 ${\le}$ $M_{\rm K} - 5\log_{10}h$ ${\le}$ -24 elliptical
galaxies have a median
star formation redshift $z_\star \sim 3$ while the median assembly redshift is 
much lower $z_{\rm assem}\le 3$. This clearly
shows that such galaxies are formed almost entirely through
the merging of pre-existing systems that formed at
higher redshift. The difference between star formation and assembly times
diminishes for lower luminosity ellipticals, suggesting fewer mergers. 
Also, the stellar population is younger in lower luminosity ellipticals compared to
their more luminous counterparts by $\sim$ 1 Gyr.
Since the other morphological types (S0-Sbc) show remarkably little
difference between their star formation and assembly redshift
distributions, it is primarily the luminous ellipticals, -25 ${\le}$ $M_{\rm K} - 5\log_{10}h$ ${\le}$ -24 whose growth is dominated by the delivery of
pre-formed stars via
mergers that happened recently, since $z~{\sim}~3$.

The star formation and assembly redshifts are virtually
identical for lenticular and spiral galaxies (S0--Scd)
which suggests that major mergers do not play an important role in the
evolution of these galaxy types. 
The  $z_\star$ distributions are strongly peaked at
low redshifts for the disk
dominated Sc-Scd galaxies. Thus, the models indicate that they should contain the youngest stellar populations with ages ${\le}$ 6 Gyrs. On the other hand, the $z_\star$ distributions
for lenticular and bulge dominated spirals become broader with decreasing
luminosity suggesting a heterogeneous group with stellar populations spanning a wide range in age.

\section{Discussion}\label{sec:Discussion}

\begin{figure}
 \includegraphics[width=80mm,viewport=5mm 15mm 190mm 265mm,clip]{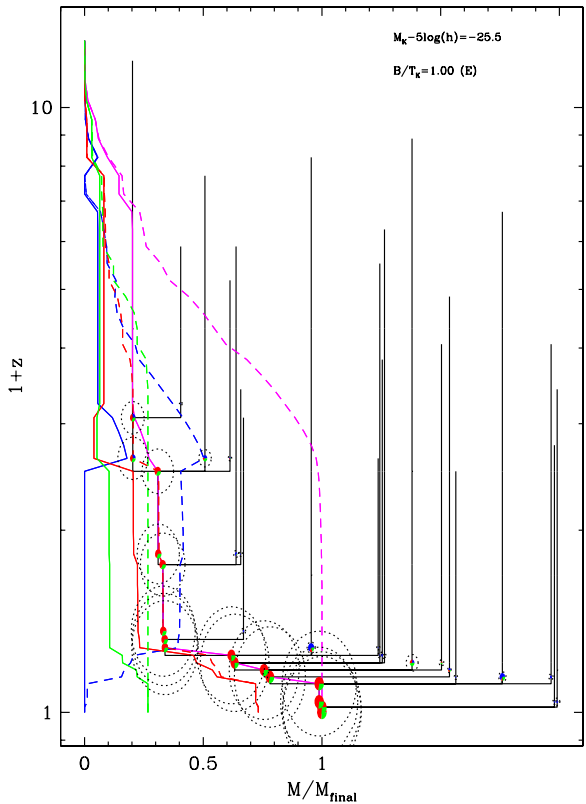}
 \caption{A formation tree diagram for a single galaxy identified at $z=0$. The K-band magnitude and bulge-to-total ratio of the galaxy at $z=0$ are shown in the upper right-hand corner. The y-axis plots redshift, while the x-axis plots stellar mass relative to that of the $z=0$ galaxy. The small pie-charts indicate the final galaxy and any progenitor galaxies (shown down to progenitors with a mass of 0.1\% of the final galaxy). Black lines connecting galaxies indicate mergers between galaxies. The radius of each pie-chart is proportional to the cube root of the stellar mass of the galaxy, while the segments of the pie chart show the relative fractions of stars in the galaxy's disk (blue), and bulge (red for stars formed by or brought in by mergers, green for stars formed or brought in by disk instabilities). The dotted circles around each galaxy are proportional to $(\Omega_{\rm b}/\Omega_0)M_{\rm h}$ and so will coincide with the size of the galaxy pie chart if a galaxy has succeeded in converting all available baryons in its halo into stars. Coloured lines indicate the total stellar mass (magenta) as well as that in the disk (blue), bulge[mergers] (red) and bulge[instabilities] (green) as a function of redshift. Solid lines show masses in the most massive progenitor, while dashed lines show the mass summed over all progenitors.}
 \label{fig:Tree}
\end{figure}

In order to understand both the trends found in \S\ref{sec:Results} and the morphological evolutionary paths of galaxies, we examine formation tree diagrams such as Fig.~\ref{fig:Tree}. These show the detailed formation history of individual galaxies and therefore provide a more detailed examination of morphological evolution. The tree shown in Fig.~\ref{fig:Tree} is for a luminous elliptical galaxy. As expected from the results of \S\ref{sec:Results} this galaxy shows significant merging activity. There is a clearly identifiable main progenitor which has been a pure elliptical since $z\approx 0.5$. Half of the stars in the final galaxy were formed by $z=3.75$ but where not assembled into a single galaxy until $z=0.26$. Approximately one quarter of the stars in the galaxy were formed as a result of a burst triggered by a disk instability event which occurred at $z\approx 5$.

\begin{figure*}
 \begin{tabular}{cc}
\vspace{-7mm} a & b \\
  \includegraphics[width=80mm,viewport=5mm 15mm 190mm 265mm,clip]{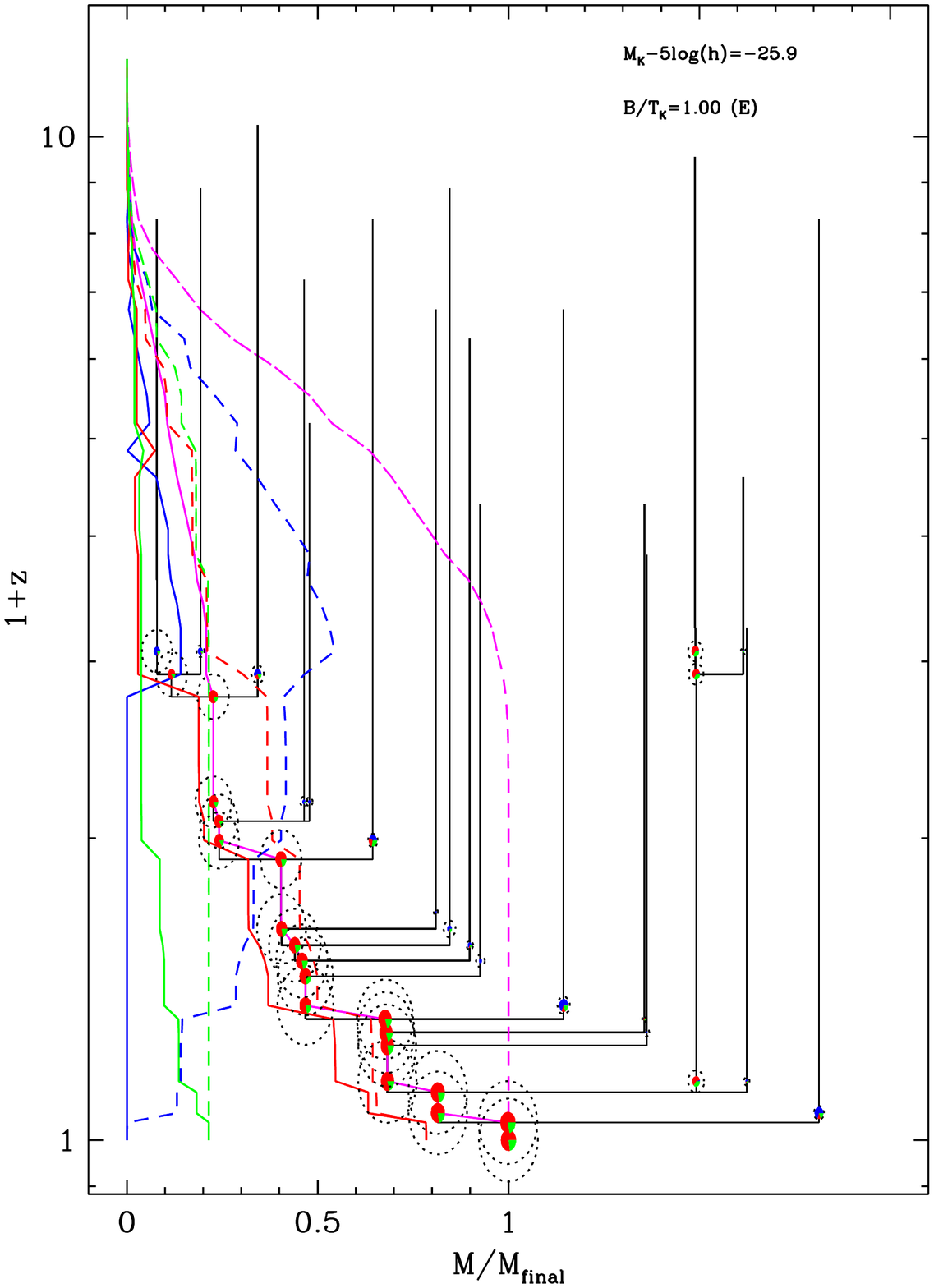} &
\vspace{7mm}   \includegraphics[width=80mm,viewport=5mm 15mm 190mm 265mm,clip]{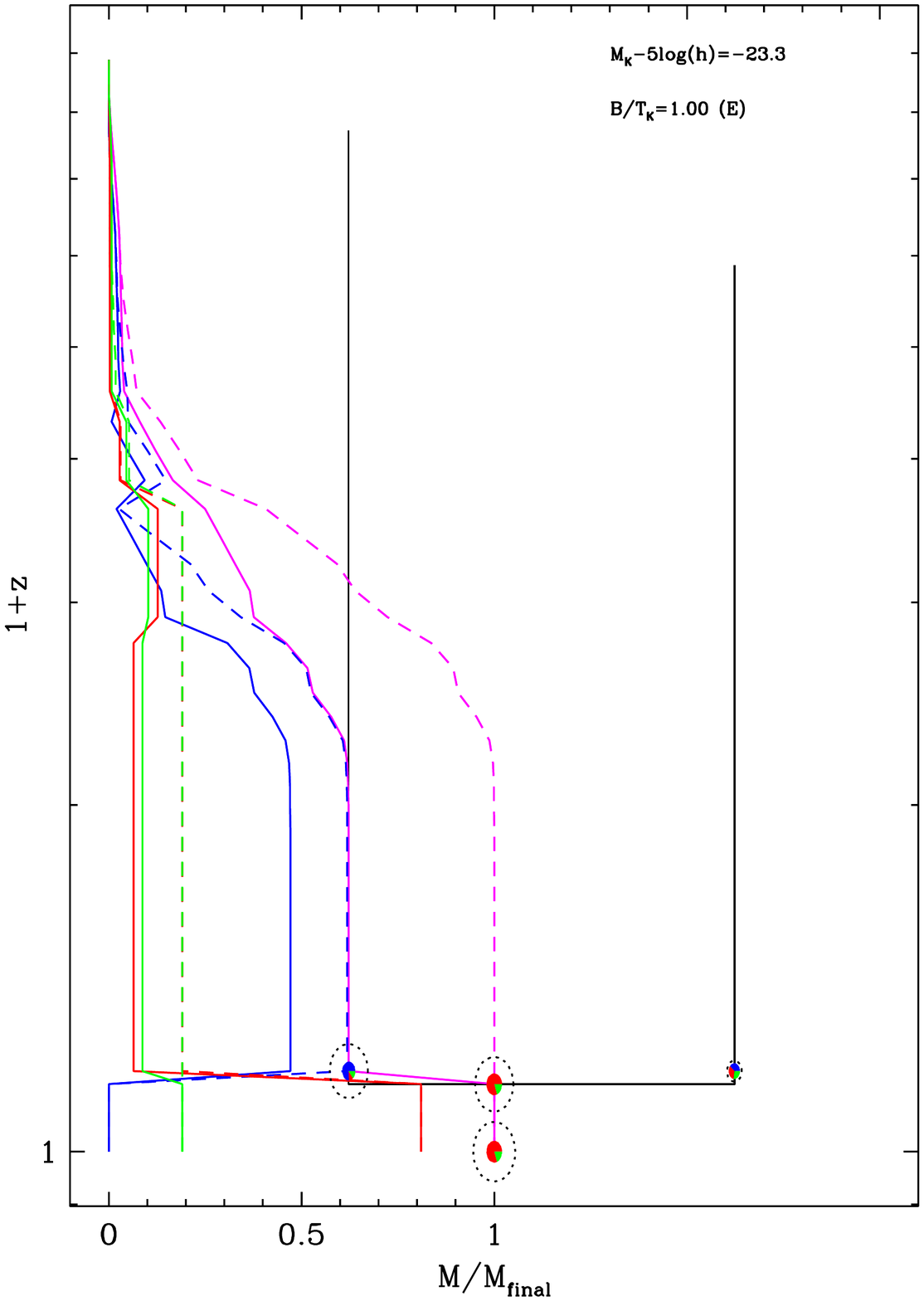} \\
\vspace{-7mm} c & d \\
 \includegraphics[width=80mm,viewport=5mm 15mm 190mm 265mm,clip]{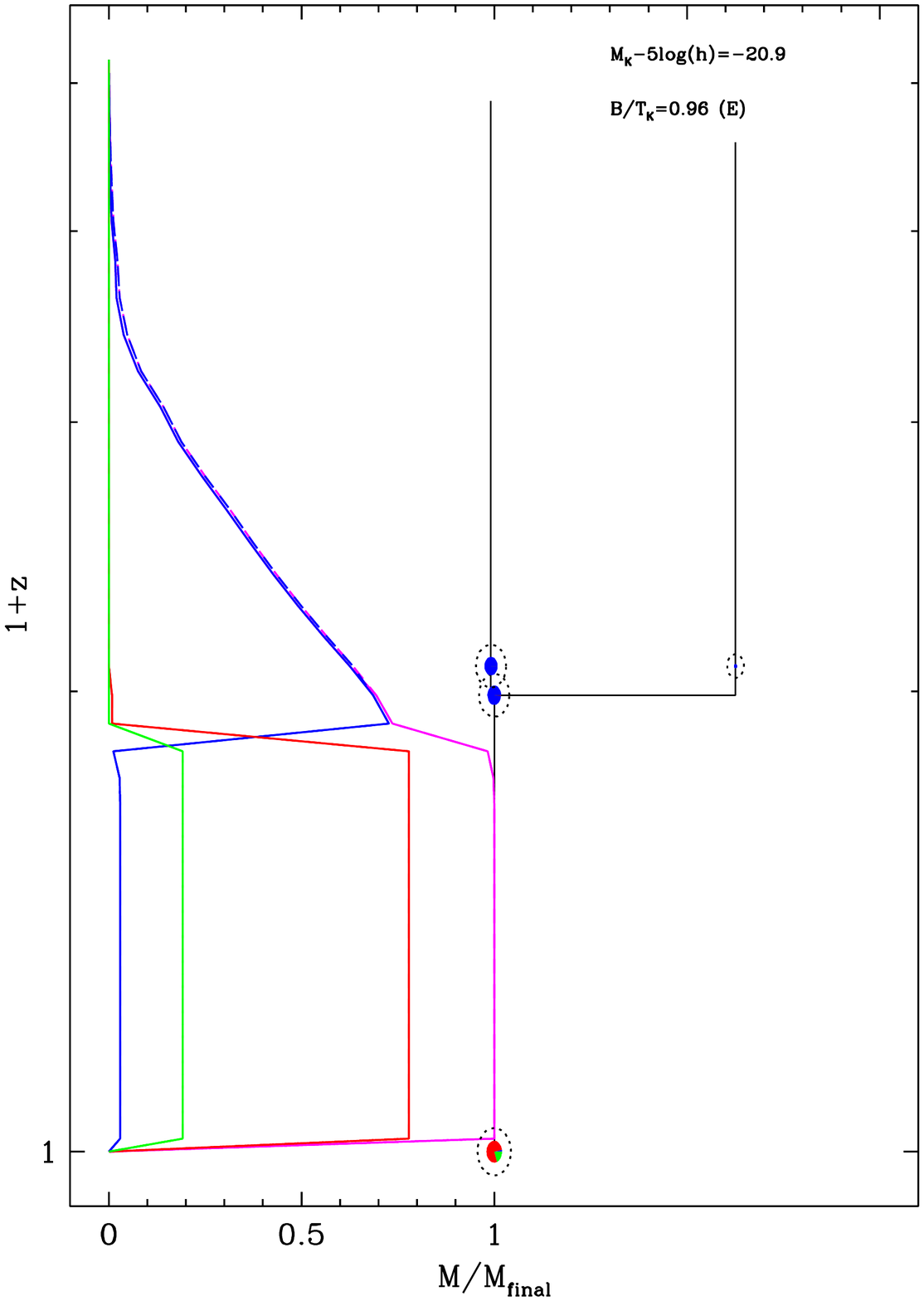} &
   \includegraphics[width=80mm,viewport=5mm 15mm 190mm 265mm,clip]{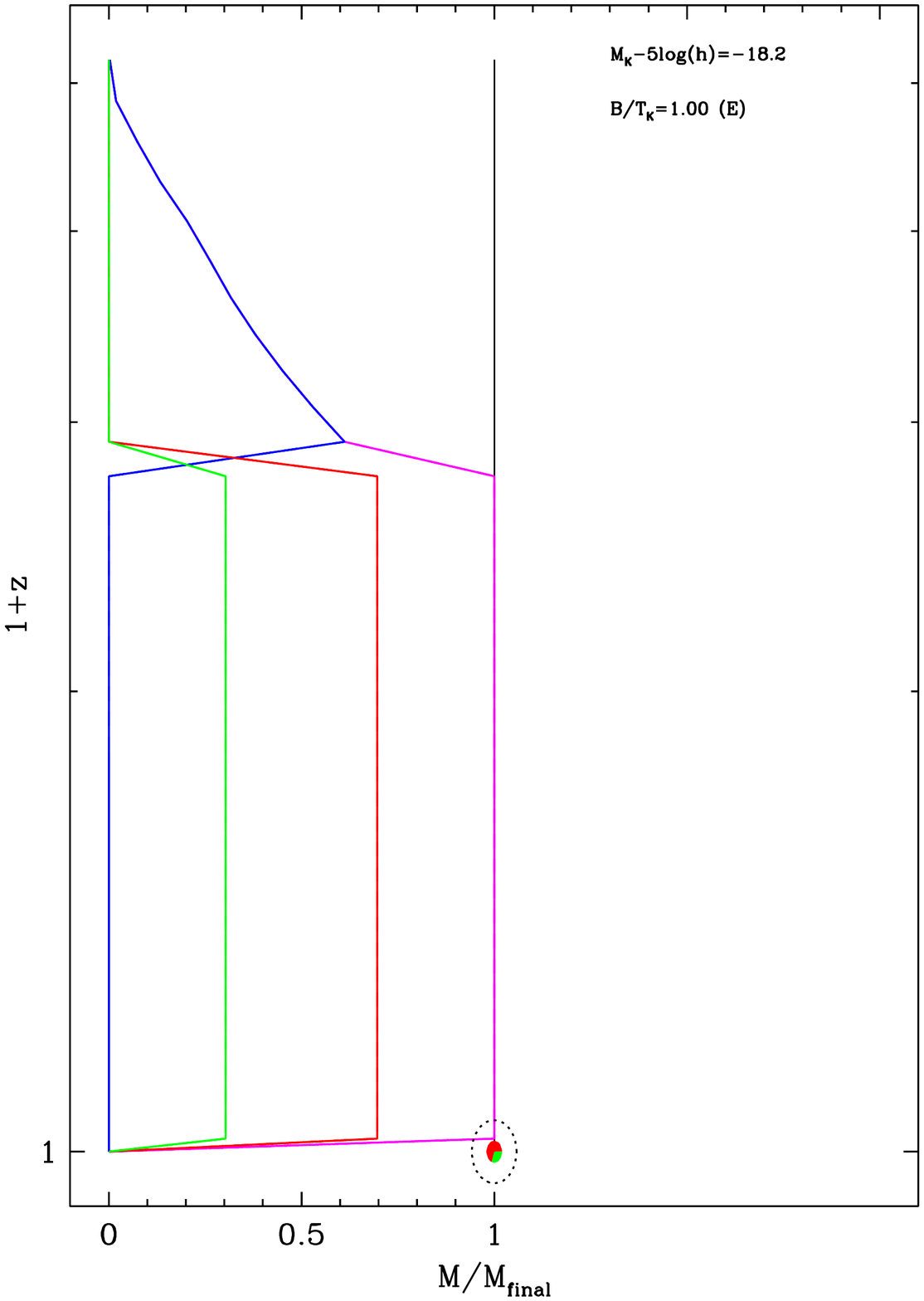} \\
 \end{tabular}
 \caption{Formation tree diagrams for elliptical galaxies shown for four different magnitudes as indicated in the panels.}
 \label{fig:TreesE}
\end{figure*}

\begin{figure*}
\begin{tabular}{cc}
\vspace{-7mm} a & b \\
\includegraphics[width=80mm,viewport=5mm 15mm 190mm 265mm,clip]{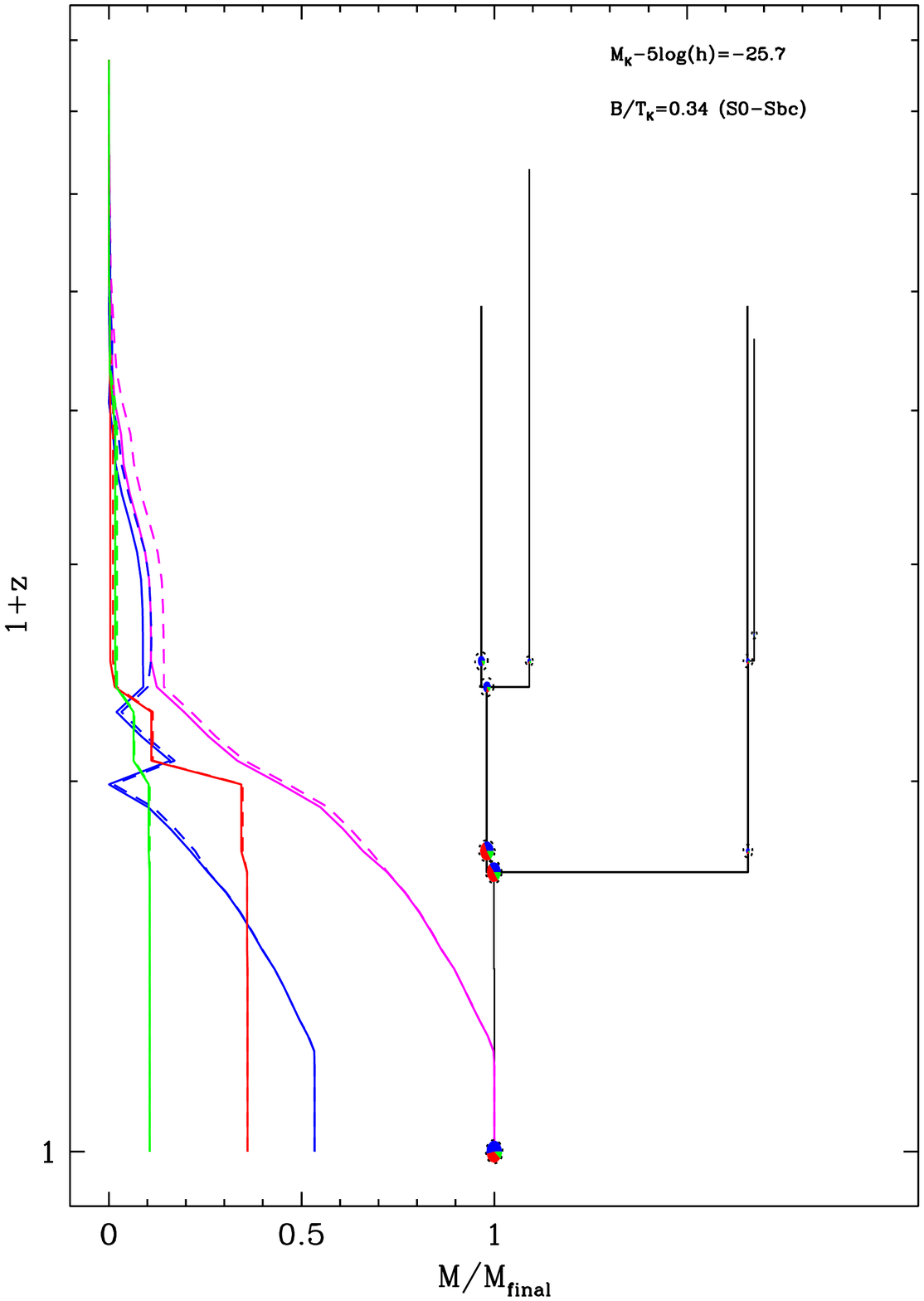} &
\vspace{7mm}   \includegraphics[width=80mm,viewport=5mm 15mm 190mm 265mm,clip]{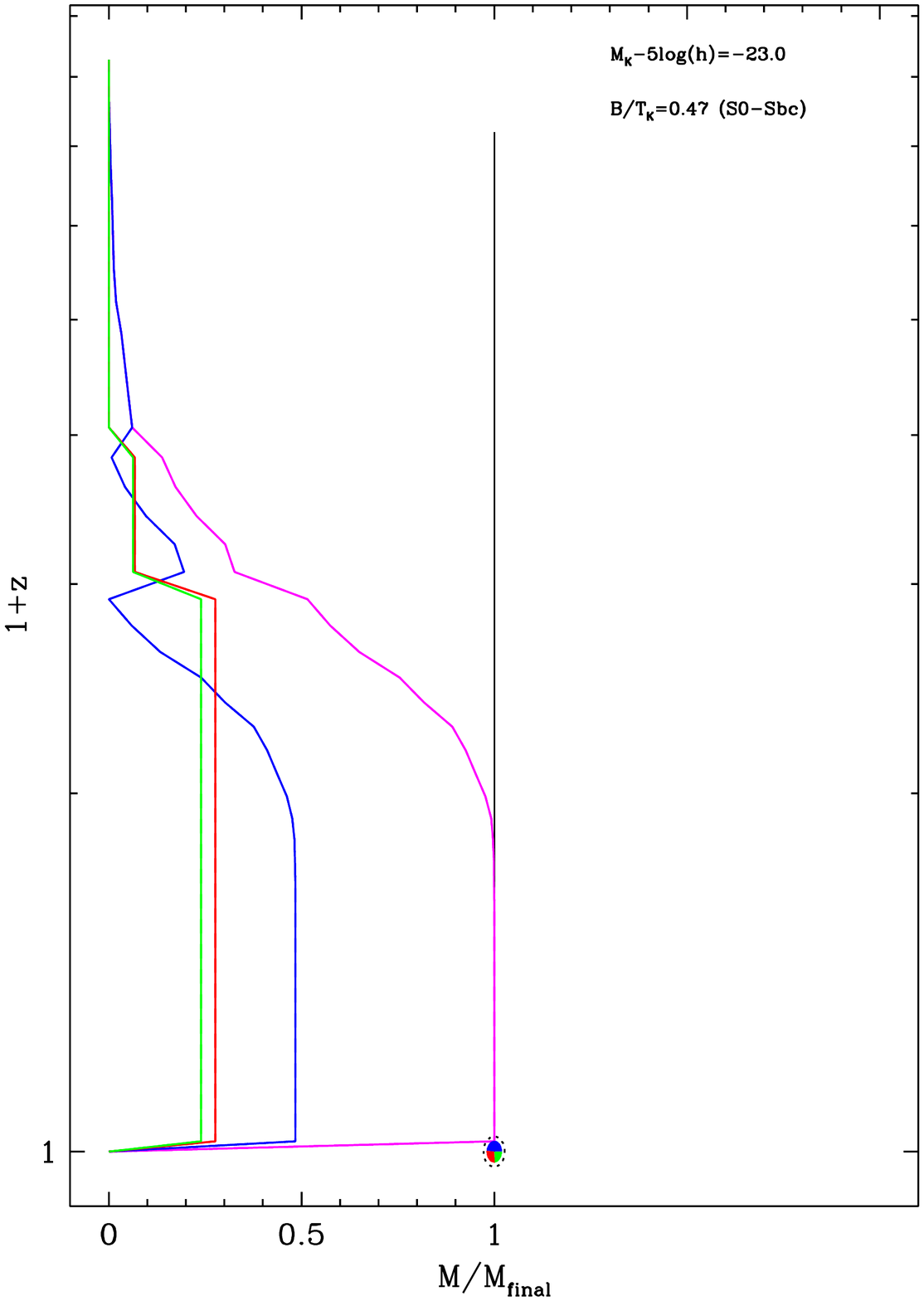} \\
\vspace{-7mm} c & d \\
 \includegraphics[width=80mm,viewport=5mm 15mm 190mm 265mm,clip]{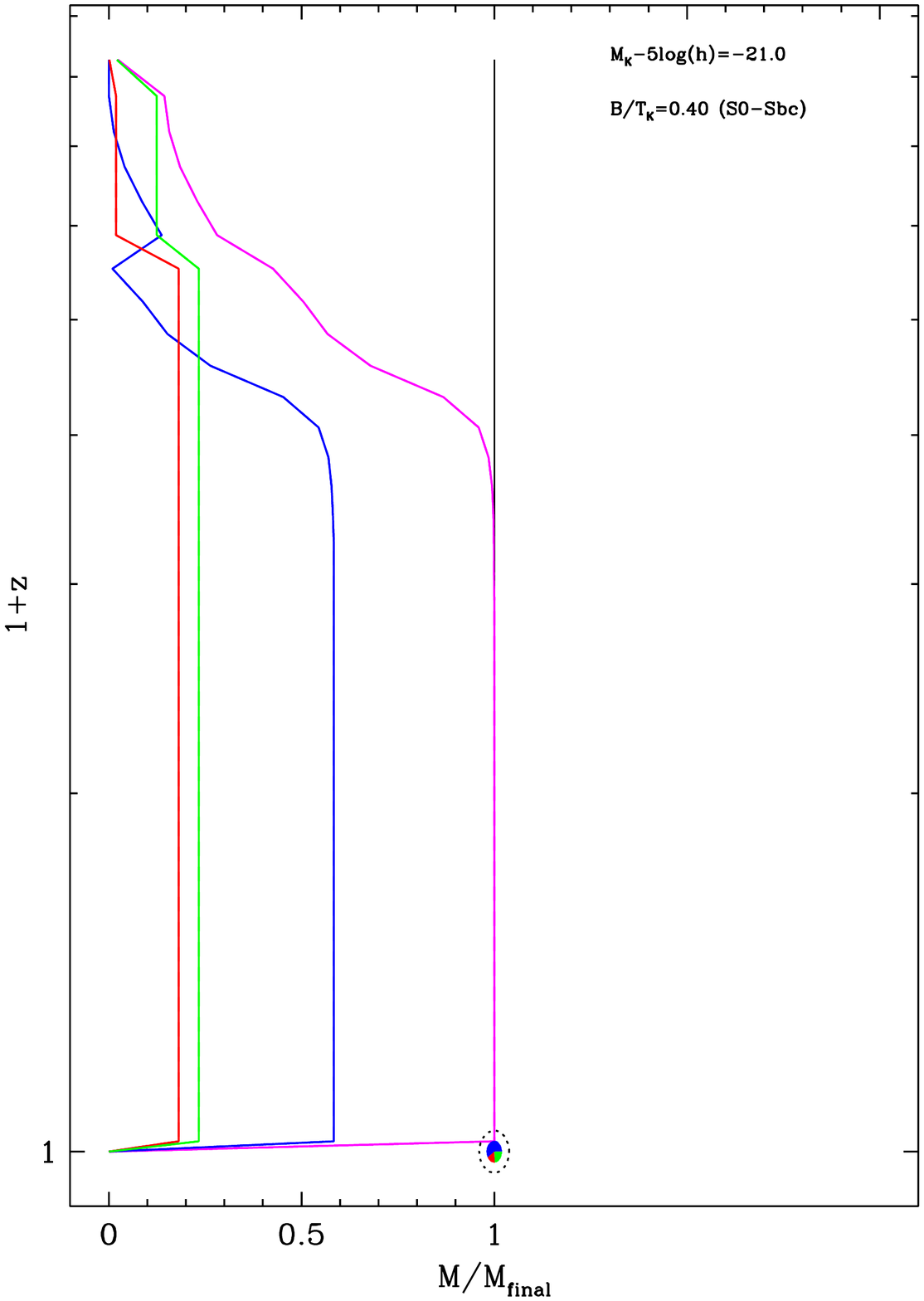} &
  \includegraphics[width=80mm,viewport=5mm 15mm 190mm 265mm,clip]{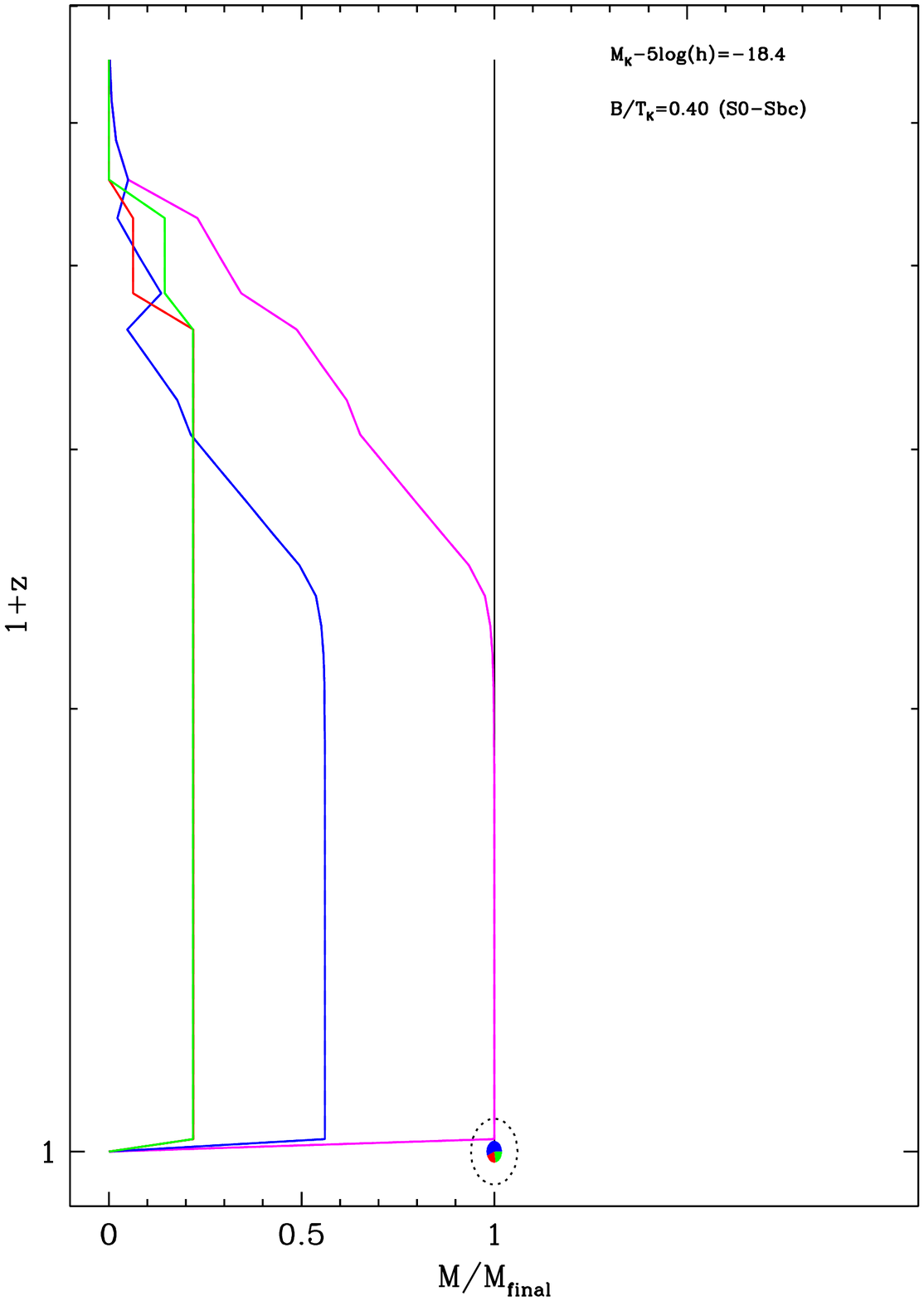} \\
\end{tabular}
\caption{Formation tree diagrams for S0-Sbc galaxies shown for four different magnitudes as indicated in the panels.}
\label{fig:TreesSa-Sab}
\end{figure*}

\begin{figure*}
 \begin{tabular}{cc}
\vspace{-7mm} a & b \\
  \includegraphics[width=80mm,viewport=5mm 15mm 190mm 265mm,clip]{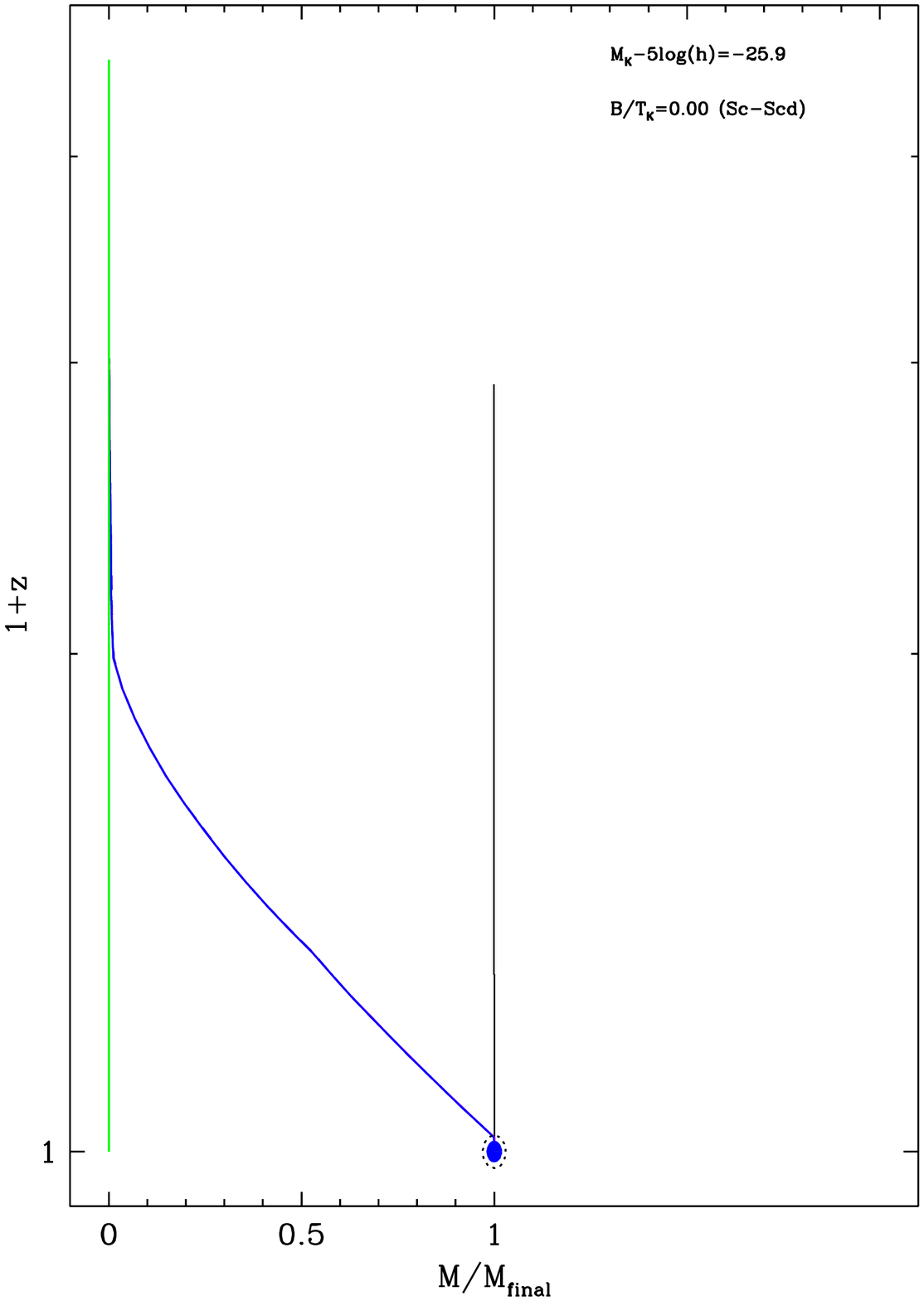} &
\vspace{7mm}   \includegraphics[width=80mm,viewport=5mm 15mm 190mm 265mm,clip]{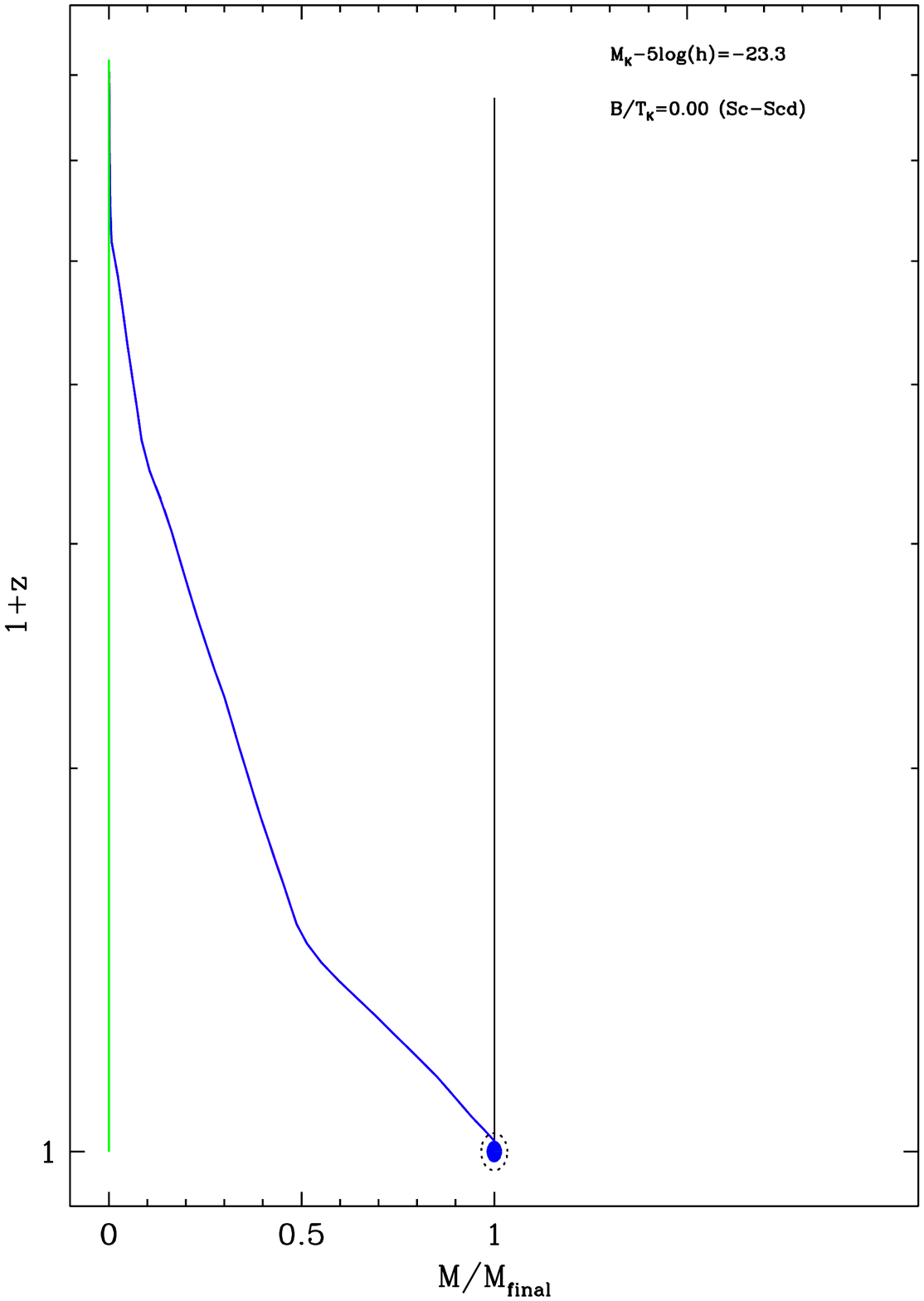} \\
\vspace{-7mm} c & d \\
   \includegraphics[width=80mm,viewport=5mm 15mm 190mm 265mm,clip]{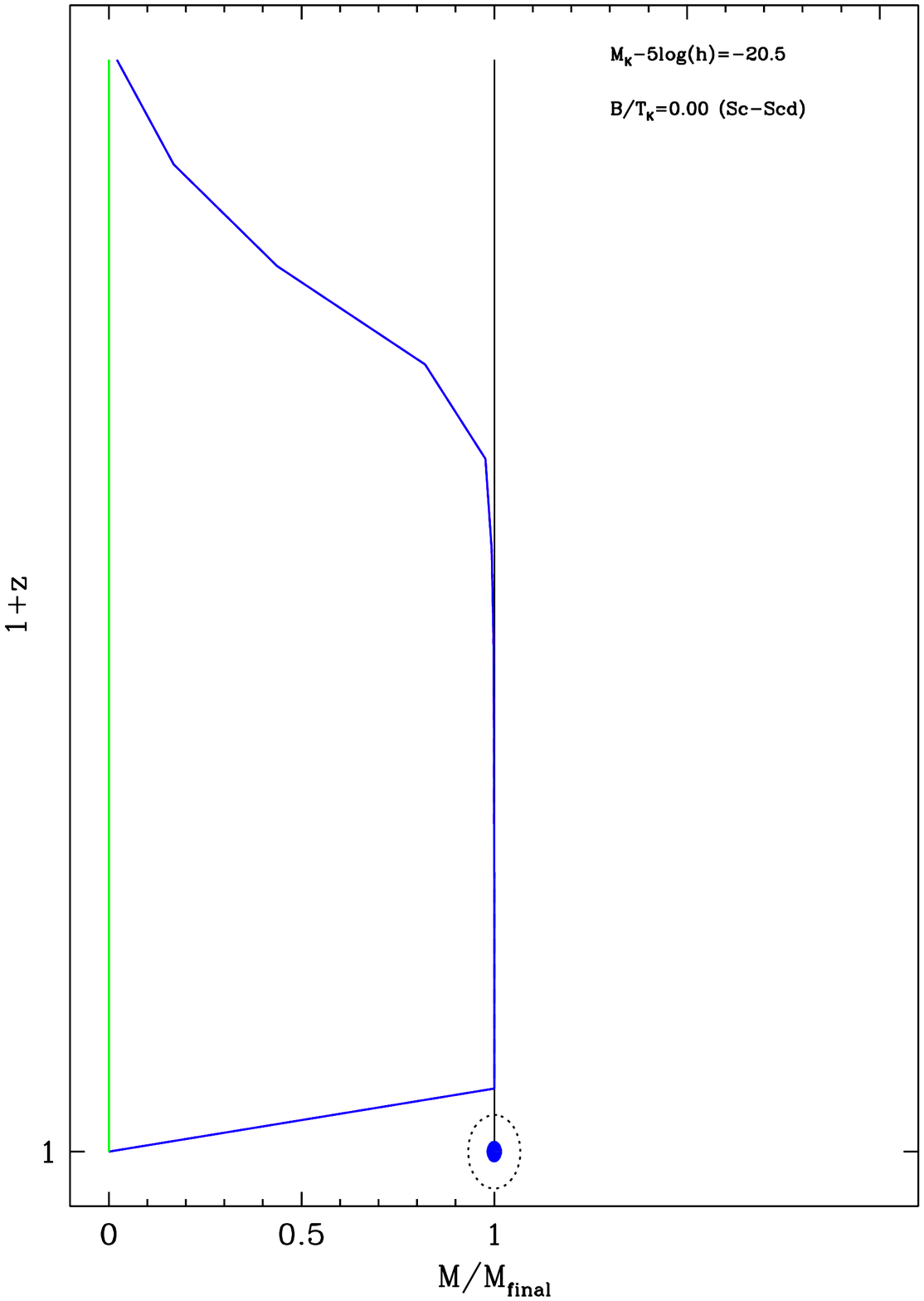} &
   \includegraphics[width=80mm,viewport=5mm 15mm 190mm 265mm,clip]{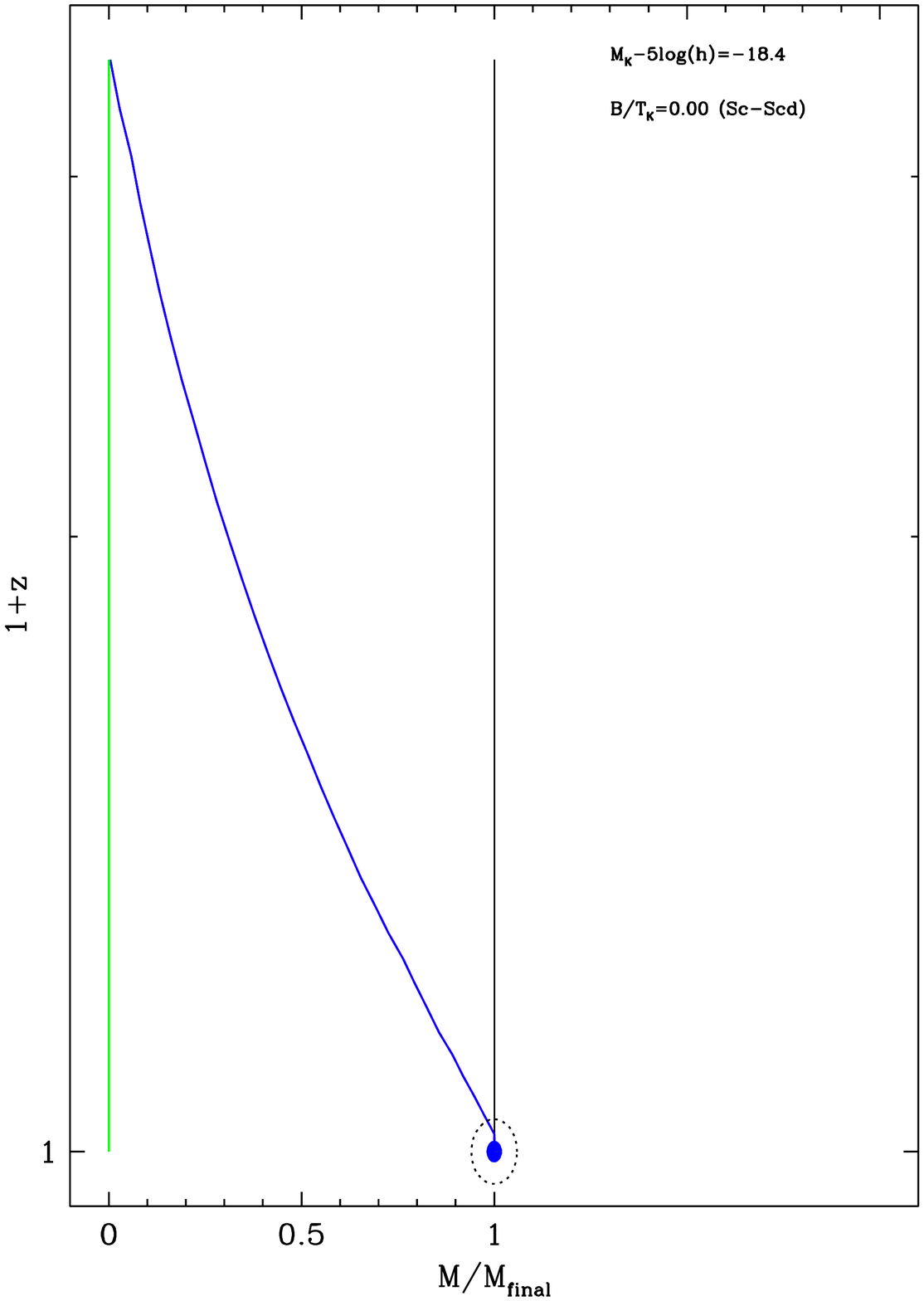} \\
 \end{tabular}
 \caption{Formation tree diagrams for Sc-Scd galaxies shown for four different magnitudes as indicated in the panels.}
 \label{fig:TreesSc-Scd}
\end{figure*}

In Figs.~\ref{fig:TreesE} through \ref{fig:TreesSc-Scd} we show tree diagrams for randomly selected galaxies in the E, S0/a-Sbc and Sc-Scd classes for a range of luminosities. (Note that the vertical axis of each plot is adjusted to stretch from $z=0$ to the earliest redshift at which the galaxy has an identifiable progenitor.) The E class shows a clear transition from forming via multiple mergers in the high-luminosity regime to forming via disk instabilities in the low luminosity regime. (A disk instabiity event in these diagrams is apparent as the solid[main progenitor] blue[disk] line drops to zero while the solid green and red lines jump to larger values.) These low luminosity ellipticals start out as disks, gradually converting their gas into stars. Eventually, the disk becomes gravitationally unstable, triggering a burst and converting all stars and gas into a spheroid. The galaxy evolves passively after this time.

It is apparent that the number of progenitors is greatly reduced for the lower luminosity galaxies\footnote{Galaxies of luminosity similar to that in panel d of this figure are affected by the resolution limit of the simulation as the halos that they occupy typically contain around 50 particles, preventing their lower mass progenitors from being resolved. For galaxies of luminosity comparable to that in panel c (and brighter galaxies) resolution is not an issue---they inhabit halos typically containing 500 particles, sufficient to permit many lower mass progenitor halos to be resolved in the simulation.}.  
This is apparent for all morphological types except for the Sc-Scd class and is a consequence of the shape of the galaxy mass function in systems of differing halo mass (e.g. \citealt{benson_effects_2003})---for the most massive halos the galaxy mass function is a very steep
function of halo mass, such that there are many more galaxies of slightly
lower
mass with which the central galaxy can merge. For lower mass halos, the
galaxy mass function is shallower, yielding relatively fewer galaxies
of lower mass. This results in fewer galaxies with which to merge and fewer
progenitors. The shape of the galaxy mass function in our
model arises because of supernovae feedback which is relatively stronger
in lower mass halos.

The S0/a-Sbc class has a much quieter history. Only the most luminous examples $M_{\rm K} - 5\log_{10}h {\le} -25$ are involved in a few significant mergers. Lower luminosity
systems experience periods of early growth characterized by multiple disk instability events  
that cause the disk mass to rise and then fall sharply. More recent growth, since $z {\sim} 2$, occurs when the disk has stabilized and is able to form stars quiescently until the gas supply is exhausted, typically between $z=1$ and $z=2$ for the lower luminosity systems and $z {\sim} 0$ for the luminous systems.

The Sc-Scd class has the most uneventful formation history of all. This is, of course, obvious from the fact that this class is defined to have almost no spheroidal component. These
galaxies must therefore have experienced no significant instability or merger events. The tree diagrams of Fig.~\ref{fig:TreesSc-Scd} show that this type of
galaxy is characterized by steady star formation in a disk fueled by cosmic infall, which continues either until $z=0$, or until the galaxy becomes a satellite and therefore loses its gas supply.

\subsection{The Merger History as a Function of Morphological Type}\label{sec:mergers}

\begin{figure}
 \includegraphics[width=75mm,viewport=5mm 55mm 205mm 245mm,clip]{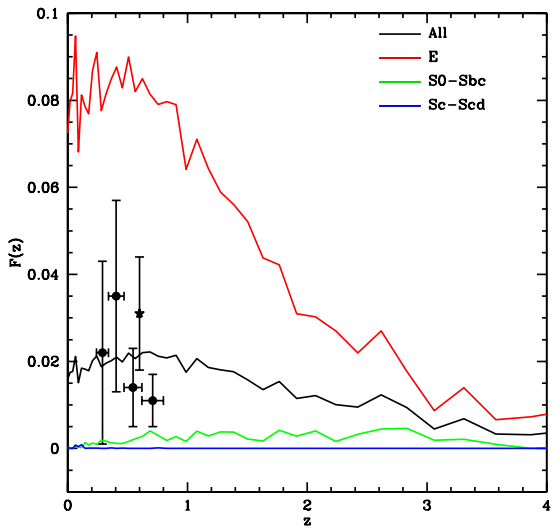}
 \caption{The fraction of galaxies with mass ${>} 2.5 \times 10^{10} M_{\odot}$ (corresponding to 
$M_{\rm K} - 5\log_{10}h {\le} -22.7$) that have undergone a major merger ($M_2/M_1 {>} 0.25$) within the past 1~Gyr. The black line shows the predicted merger fraction averaged over all morphological types, while the blue, green and red lines show the predicted merger fraction for disk, disk+bulge and bulge morphological classes respectively. Points show estimates of the merger fraction based on observations described by \protect\cite{jogee_history_2009} and \protect\cite{lopez-sanjuan_robust_2009} (circles and star respectively).}
 \label{fig:MergerRateJogee}
\end{figure}

Figure~\ref{fig:MergerRateJogee} illustrates the merger history as a function of morphological type. 
$F(z)$ is the fraction of galaxies with mass ${>} 2.5 \times 10^{10} M_{\odot}$ (corresponding to 
$M_{\rm K} - 5\log_{10}h {\le} -22.7$)
that have undergone a major merger ($M_2/M_1 {>} 0.25$) within the past 1Gyr. These selection criteria allow a direct comparison with the 
observationally determined rates of  \cite{jogee_history_2009} and \cite{lopez-sanjuan_robust_2009}. 
Galaxies are segregated according to their $z=0$ morphology
and the reported merger fractions resulted from integrating their respective merger trees over $z$. The model predicts an average merger fraction
integrated over all galaxy types of ${\sim}$ 2\% which compares favorably with the observed lower limit to the major merger fraction for the sample S1 of  \cite{jogee_history_2009} and 
the merger fraction measured at $z = 0.6$ for a slightly more luminous sample by \cite{lopez-sanjuan_robust_2009}.
The model reveals that the merger rate is highest for elliptical galaxies; ${\sim}$ 8\% of ellipticals at $z = 0$ resulted from a major merger that occurred in the past 1 Gyr. The model elliptical merger rate is significantly higher than the merger rate estimated by \cite{masjedi_growth_2008} for 
a sample of luminous red galaxies in close pairs. The model elliptical merger rate predicts 
significant evolution of the elliptical LF, as discussed in more detail in the next section.
Very few of the
$z=0$ lenticular and (S0/a-Sbc) spiral galaxies resulted from a major merger and none of the Sc-Scd disk galaxies. Thus, the model predicts that 
Sc-Scd galaxies exist in the local universe because they have not been involved in any collisions.

We compare our predicted merger rates with other recent attempts to measure
this quantity from either N-body simulations of dark matter or from semi-analytic models of galaxy formation. Before doing so, we remind the reader that
our merger rate is that of galaxies, and so there is no reason to expect it to
coincide with rates of, for example, dark matter halo merging or dark matter
subhalo destruction (both of which are commonly measured quantities). In
particular, \cite{stewart_galaxy_2009} examined the rates of both halo merging
(when a halo first crosses the virial radius of a larger halo, thereby becoming
a subhalo) and subhalo destruction (when a subhalo loses 90\% of its mass) in
an N-body simulation of dark matter. \cite{stewart_galaxy_2009} find merger
rates which increase monotonically with increasing redshift, in disagreement
with our own findings. However, we caution that these results are based purely
on dark matter properties (with a simple, redshift dependent mapping of galaxy
luminosity to dark matter halo mass) and so do not measure the same quantity
as presented in this work. Similar results were found by \cite{genel_halo_2009}.
On the other hand, \cite{bertone_comparison_2009}  find merger fractions
comparable to ours, with a broad peak around $z=1.5$ and a gradual
decline towards higher redshifts. The agreement is not surprising, however, as 
they derive merger rates from a semi-analytic model
similar to the one used here.

\subsection{The Evolution of the Elliptical Galaxy Luminosity Function}

\begin{table*}
\begin{center}
\caption{$K$-band Luminosity Function Fit Parameters}
\label{tb:LFParam}
\begin{tabular}{ccccc}
\hline
{\bf Sample} & {\bf $\phi_{*}$/${\it h}$$^{3}$} & {\bf M$_{*}$$-$ 5log$_{10}$${\it h}$} & {\bf $\alpha$}  & {\bf j} \\
 & {\bf 10$^{-4}$  galaxies\,Mpc$^{-3}$ mag$^{-1}$} &  {\bf mag} &  & {\bf 10$^{7}$${\it h}$L$_{\odot}$\,Mpc$^{-3}$}   \\
 \hline
z = 0 &  16.1  ${\pm}$ 0.5  & $-$23.93 ${\pm}$ 0.07 & $-$0.48 ${\pm}$ 0.02 &  10.4  \\
z = 1 & 7.1  ${\pm}$ 0.3   & $-$23.77 ${\pm}$ 0.09 & $-$0.70 ${\pm}$ 0.02 & 4.3  \\
z = 2 &  4.9  ${\pm}$ 0.2   & $-$23.15 ${\pm}$ 0.08 & $-$0.82 ${\pm}$ 0.01 & 1.9  \\
\hline
\end{tabular}
\end{center}
\end{table*}

\begin{figure}
 \includegraphics[width=80mm]{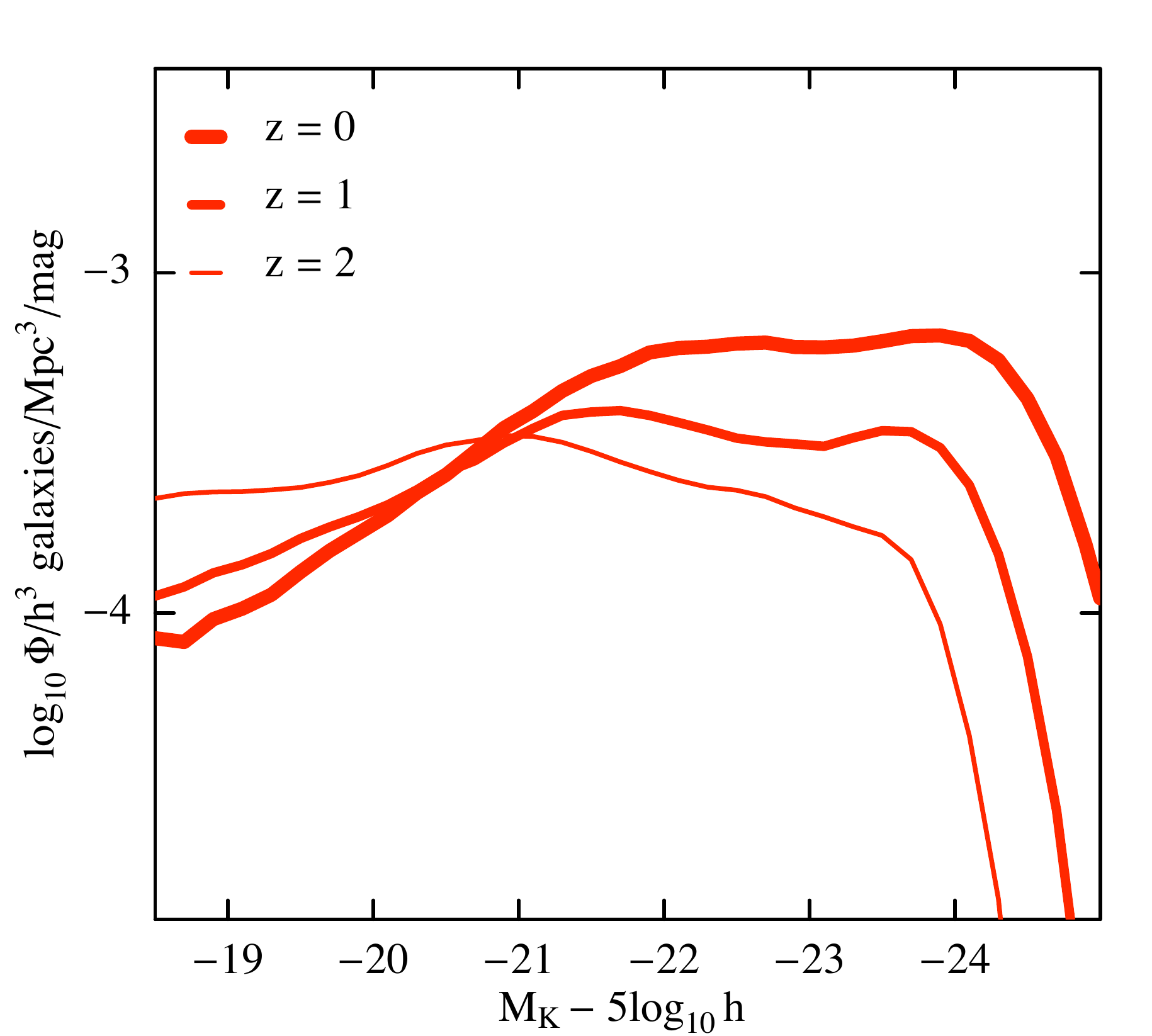}
 \caption{The predicted evolution of the elliptical galaxy luminosity function with redshift. The
 luminosity functions have been k-corrected and passively evolved to $z=0$. So, the differences
 reflect the mass build up, due primarily to mergers at the high luminosity end: -24.5 ${\le}$ M$_{K}$$-$ 5log$_{10}$${\it h}$ ${\le}$ -25.5 }
 \label{fig:EllipticalEvolution}
\end{figure}

The model predicts significant evolution of the elliptical galaxy K-band luminosity function as illustrated in Figure~\ref{fig:EllipticalEvolution}.
The LF at each redshift has been k-corrected and further corrected for passive evolution to $z = 0$ so that the LFs may be compared on an equal basis. Within the {\sc Galform} model, k-corrections and passive evolution can be computed precisely. The model predicts the full SED of each galaxy and so we can simply shift the filter to the galaxy rest frame and compute the absolute magnitude which will then automatically include the k-correction. For passive evolution, the model predicts the mix of stars in each galaxy at each redshift, including their distribution of ages and metallicities. To compute the magnitude of the galaxy at $z=0$ assuming passive evolution (i.e. no further star formation) we simply artificially increase the ages of these stars by the lookback time to the current redshift and then sum their SEDs to give the net SED of the galaxy passively evolved to $z=0$. The passively evolved magnitude of the galaxy is then trivially found by integrating under the appropriate filter response function. Our k and evolution corrections are, at least at low redshifts, in reasonable agreement with those determined observationally by \cite{bell_optical_2003}. For example, in the K-band at $z=0.5$ we find k and evolution corrections of approximately $k=-0.75$ and $e=0.55$, to be compared to $k=-1.05\pm0.15$ and $e=0.4$ from \cite{bell_optical_2003}. 

Figure~\ref{fig:EllipticalEvolution} shows that the space density goes up at the high luminosity end and  down at the low luminosity end as the redshift decreases.  These evolutionary changes have been
quantified in Table~\ref{tb:LFParam} in terms of the Schechter parameters $\phi_{*}/h^{3}$,   $M_{*} - 5\log_{10}h$ and $\alpha$. The model results indicate that the mass build-up of ellipticals is 
reflected in the $K$-band mostly through the large ${\sim}$ 226\% increase in $\phi_{*}$/${\it h}$$^{3}$ since $ z = 1$, and a small, $0.2 {\pm} 0.1$ mag, brightening of M$_{*}- 5\log_{10} h$. 

Parameterizing the LFs allows the luminosity density {\it j} 
to be calculated by integration,

\begin{equation}
j =  \int \phi (M) 10^{0.4(M_\odot - M)} dM
\end{equation}

where M${_\odot}$ is the absolute magnitude of the Sun, corresponding to 3.32\,mag at $K$ \citep{bell_optical_2003}. The luminosity density is important as it can provide a constraint on the mass density of stars
in galaxies, given a mass to light ratio. 
The total $K$-band luminosity density at each redshift was calculated by integrating equation 3 over the
interval $-$25 ${\le }$ M$_{K}$$-$5log$_{10}{\it h}$ ${\le}$$-$19\,mag and the results are summarized
in Table~\ref{tb:LFParam}.
These values 
indicate that the luminosity density, or equivalently, the mass density has increased by about a factor of 2.4 since $z = 1$. The predicted evolution since $z = 2$ is even more dramatic. These model results concur in general with recent observational results 
suggesting that the stellar mass contained within the red galaxy population has at least doubled since 
$z ~{\sim}$ 1 \citep{bell_nearly_2004, willmer_deep_2006, blanton_galaxies_2006, faber_galaxy_2007,
brown_evolving_2007}. 

Our model results further indicate that the mass density of luminous ellipticals grows at the expense of their lower luminosity counterparts causing the faint end slope to turn downwards with decreasing
redshift. Such low luminosity ellipticals are presumably satellites of their more luminous counterparts, they are difficult to detect observationally with current technology and contribute negligibly to the 
luminosity density evolution of ellliptical galaxies. At the other end of the luminosity range, recent observations  \citep{brown_evolving_2007, cool_luminosity_2008} 
indicate that the luminous M$_{B}$$-$ 5log$_{10}$${\it h}$ ${\le}$ -21 
red galaxy population, loosely identified with ellipticals, were already in place at $z = 1$ whereas the 
corresponding model elliptical luminosity density, integrated between  -24.5 ${\le}$ M$_{K}$$-$ 5log$_{10}$${\it h}$ ${\le}$ -25.5, shows an increase by a factor of 4 since $z = 1$. Thus, the model predicts significant evolution of the most luminous elliptical galaxies since $z = 1$ that observers do not see. 
One possible reason for the discrepancy is that
the colour selection criteria employed by observers may result in significant contamination of the observational datasets with galaxies that are not actually ellipticals, causing elliptical galaxy evolution to be underestimated. This potential 
problem with the observations was noted previously by \cite{faber_galaxy_2007}. It is crucial
therefore to understand the connection between the luminous red galaxy population seen at  $z~{\sim}$ 1 and the elliptical population at $z = 0$. Are the luminous red galaxies seen at $z~{\sim}$ 1 the progenitors of todays ellipticals
or are they dusty objects destined to become what we recognise in the local universe
as lenticulars, or early-type spiral galaxies?
Of course, another possibility is that the model may be at fault in overestimating the evolution of luminous ellipticals although, in defense of the model, it does succeed in predicting the observed space density of ellipticals at $z = 0$\footnote{Since the parameters of the model were adjusted to match the observed morphologically segregated luminosity functions this is not strictly a prediction. However, the unadjusted \protect\cite{bower_breakinghierarchy_2006} model also yields a similar space density for elliptical galaxies as shown in Figure~\ref{fig:BestLF}.} and the observed type averaged merger rate ( \S\ref{sec:mergers}) and the observed star formation rate, as described in more detail in the next section.

\subsection{The Star Formation History as a Function of Morphological Type}

With regard to elliptical galaxies, the results presented in Figure~\ref{fig:StarsAssem} confirm the trend seen observationally (e.g., \cite{thomas_epochs_2005} and references therein) that the star formation redshift for elliptical galaxies is correlated with their mass (luminosity) in the sense that the stars comprising the more massive (luminous)  ellipticals are old.
The model predicts that half the stars comprising todays luminous, $-25 {\le}$ $M_{\rm K} - 5\log_{10}h$ ${\le} -24$, ellipticals formed between redshifts of 2 and 4 with 
ages ${\sim}$ 10~Gyr and were subsequently assembled via mergers, that occurred since $z~{\sim} 3$, into the objects 
we recognise in the local universe as luminous elliptical galaxies. In contrast, stars comprising the lower luminosity, $-24 {\le} M_{\rm K} - 5\log_{10}h$,  ellipticals 
are younger with ages spanning 4--10 Gyrs and that were created in star formation episodes the most recent of which were triggered by disk instabilities that occurred
since $z\sim 2$. Thus, the seemingly anti-hierarchical behavior depicted in Figure~\ref{fig:StarsAssem} results from the fact that there are different mechanisms governing the star formation history of ellipticals: mergers for the luminous ones and disk instabilities for the less luminous ones. These results on formation redshift are insensitive to merger tree mass resolution (with galaxy mean stellar ages changing by only a few percent if we increase the merger tree resolution by a factor of 16) and to the details of the treatment of disk instabilities (adopting the minimalistic model of Appendix~\ref{app:altDiskInstab} results in changes to mean stellar age of less than 10\%). We note that, in our model, the amount of mass brought in to low luminosity ellipticals by minor mergers is relatively small due to the strong supernovae feedback which acts to suppress galaxy growth in lower mass halos.

\begin{figure}
 \includegraphics[width=80mm]{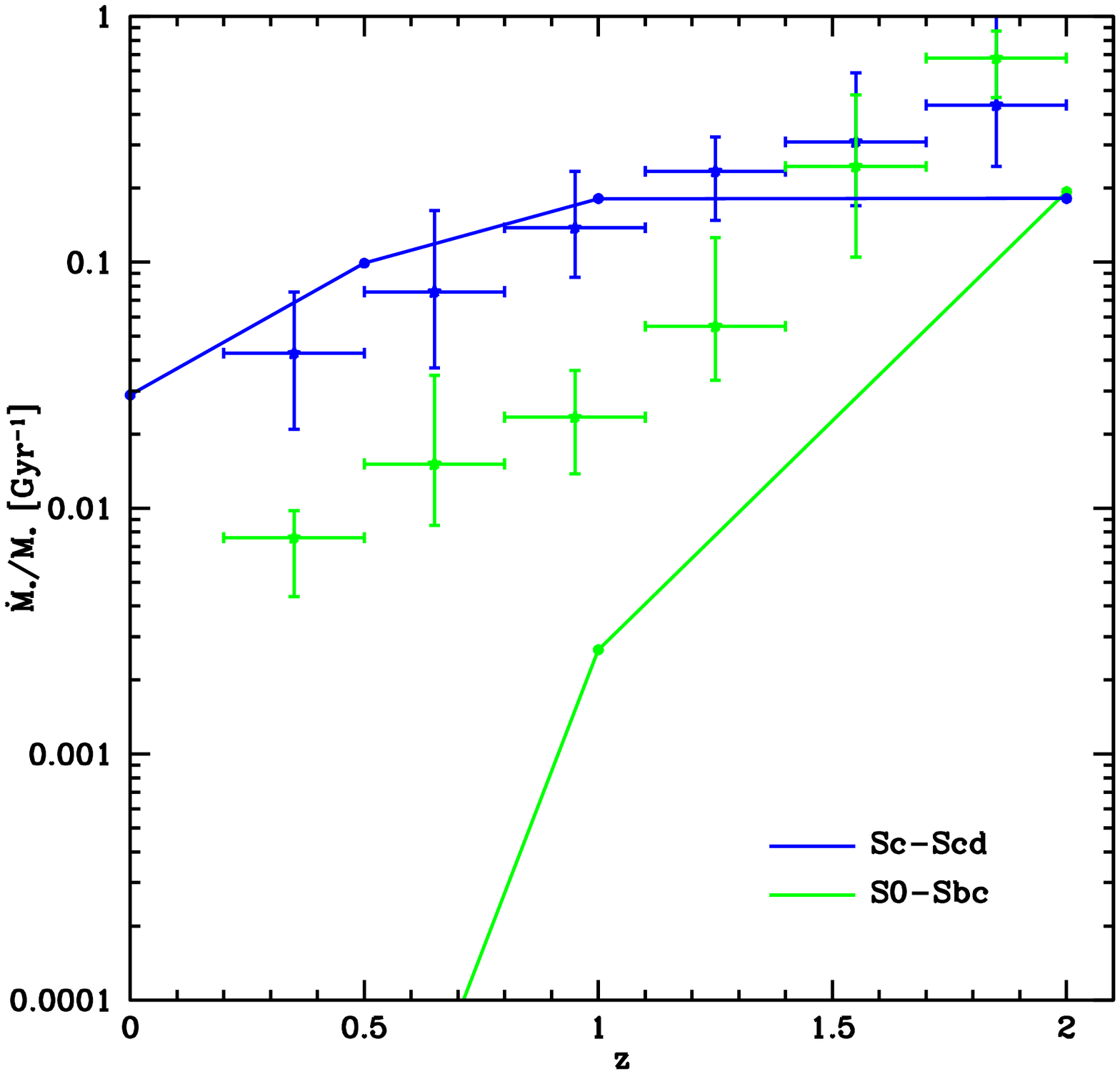}
 \caption{The evolution of the specific star formation rate with redshift for massive ${\ge}$ 10$^{11}$ M$_{\odot}$ galaxies. The blue and red crosses identify the disky galaxies  and spheroids  of \citet{prez-gonzlez_exploringevolutionary_2008}. The blue and green lines identify the model  disk and disk+bulge galaxies respectively. The model SSF rates for the bulge dominated elliptical galaxies are essentially zero.}
 \label{fig:SSFEvolution}
\end{figure}

In Figure~\ref{fig:SSFEvolution} we compare model predictions concerning
the evolution of the specific star formation (SSF) rates with those measured 
recently by \citet{prez-gonzlez_exploringevolutionary_2008} for massive ${\ge}$ 10$^{11}$ M$_{\odot}$ galaxies. The measured star formation rates are computed assuming a Chabrier initial mass function and so correspond to the total star formation rate (i.e. including stars of all masses). The same is predicted by the model (in which we compute the total star formation rate from the known dynamical time and gas content of each model galaxy). The principal merits of the \citet{prez-gonzlez_exploringevolutionary_2008}
work are that the SSF rates are based on UV to Mid IR photometry and the results are segregated by
visual morphology. The figure shows that the observed SSF rates for `disky galaxies' and `spheroids'
are bounded by the model predictions for `disk' and `bulge+disk' galaxies and both the model
and observations show a well established trend  that the SSFR is decreasing with 
decreasing redshift, with implications for the stellar mass accumulated in these types of systems \citep{dahlen_evolution_2007, chen_constraintsstar_2009}.

\subsection{The Secular Evolution of Bulges}

The model predicts that the bulges in spiral galaxies have grown through a combination of disk instabilities and minor mergers.  Observationally, the colour of bulges 
are more similar to the disks in which they reside than to other bulges \pcite{balcells_colors_1994,peletier_ages_1996}. This coupling between bulge and disk colour has been observed
out to $z {\sim} 1$ \pcite{dominguez-palmero_bulges_2008,dominguez-palmero_nature_2009} where younger, bluer high surface brightness bulges have been observed, suggesting an epoch of bulge
formation.  In concurrence with the observations, the model predicts that  bulges in luminous $M_{\rm K} - 5\log_{10}h {\sim} -23$ spiral galaxies have grown 
significantly through a combination of disk instabilities and minor mergers since $z {\sim} 2$  and the model further predicts that bulges have been
growing for a much longer time period, since $z {\sim} 4$, in lower luminosity systems. 

\section{Conclusions}

The {\sc Galform} semi-analytic model of galaxy formation \citep{bower_breakinghierarchy_2006} has been employed to reproduce and explain trends in galaxy morphology seen across a wide range of luminosities.
Small adjustments to the model parameters permit excellent agreement with the observed morphologically selected $K$-band LF of \cite{devereux_nearby_2009}, although the unmodified \cite{bower_breakinghierarchy_2006} model also performs very well. While predicting morphology from semi-analytic models (and numerical simulations) remains challenging, a conservative approach has been adopted by categorising model galaxies into just three broad classes: E (``pure spheroid''), S0-Sbc (``intermediate'') and Sc-Scd (``pure disk'') based on their $K$-band bulge-to-total luminosity ratio. Within this conservative categorisation, the results presented are converged with respect to the resolution of the N-body-derived merger trees (see Appendix~\ref{app:ResStudy}).  Our conclusions follow.

\begin{itemize}
 \item The Sc-Scd disk galaxies, with ${\sim}$ 10\% or less of their K-band light coming from the bulge, can form in abundance even in a hierarchical cold dark matter cosmology in which merging is, in general, commonplace. The median merger history of the halos that these galaxies inhabit is not significantly different from that of elliptical galaxies which do experience significant merging. However, the variation around that median history is much less for the Sc-Scd galaxies than for ellipticals. The Sc-Scd class typically experience a very simple accretion history, deviating little from the median and exhibiting relatively few large merging events. The formation history of these systems is therefore determined by the rate of growth of their dark matter halo and the corresponding rate of gas supply, cooling and infall. Galaxies in this class have typically been forming stars in their disks over large fractions of cosmic history and continue to do so at the present day. \cite{steinmetz_2002} and, more recently, \cite{dutton_origin_2009} have suggested that producing bulge-less disk galaxies in sufficient numbers could be challenging in a CDM universe. In fact, because there is a sufficiently large number of dark matter halos with relatively quiet merging histories, bulge-less disk galaxies
 can prevail, in the numbers observed. Additionally, strong feedback prevents the formation of relatively massive satellites around these galaxies which are predominantly low mass.
\item Luminous elliptical galaxies form through multiple mergers of smaller progenitors. For the most massive ellipticals these mergers tend to be ``dry'' and so contribute no new star formation. 
Interestingly, the {\sc Galform} model does not produce the dwarf elliptical sequence which should emerge at $M_{\rm K}-5\log_{10}h \sim -21$ mag. Thus, the model prediction concerning the shape of the LF for  low luminosity ellipticals (with $M_{\rm K}-5\log_{10}h> -21$ mag) and the model 
prediction that they form through disk instability events is uncertain.
\item The intermediate morphological types, S0-Sbc, are characterized by an early period of merging and disk instability events which form the bulge, followed by an extended period of steady, uninterrupted disk growth.
\item The model results concur with several observational results including the type averaged galaxy merger rate, measurements of the specific star formation rate, the secular evolution of spiral bulges,
the seemingly anti-heirarchical evolution of elliptical galaxies and the factor of ${\sim}$ 2 growth
in the luminosity density of elliptical galaxies since z ${\sim}$ 1. 
With regard to the last point however, the model predicts significant evolution at the high luminosity end of the $K$-band elliptical galaxy LF that has yet to be confirmed observationally.
\end{itemize}

\section*{Acknowledgments}

AJB acknowledges support from the Gordon and Betty Moore Foundation and would like to acknowledge the hospitality of the KITP at the University of California, Santa Barbara where part of this work was completed. This research was supported in part by the National Science Foundation under Grant No. NSF PHY05-51164.

\bibliographystyle{mn2e}
\bibliography{KLF_Morphology}

\appendix

\section{Robustness of Model Morphologies}

In this appendix we explore to effects which could influence the morphological properties of model galaxies, and demonstrate that, in fact, our results are robust to these effects.

\subsection{Resolution Study}\label{app:ResStudy}

Since morphological evolution in our model is driven at least partly by merging activity, it is important to assess whether the resolution of the N-body merger trees that we utilize causes us to miss some lower mass merging activity and, therefore, to incorrectly estimate morphologies of some galaxies. While we currently do not have suitable N-body merger trees with higher resolution available to us for a resolution study, we can instead utilize merger trees constructed using a modified extended Press-Schechter approach (see \citealt{parkinson_generating_2008}) which is designed to produce trees which are statstically equivalent to those found in N-body simulations. The advantage of this approach is that it allows us to increase the resolution to almost arbitrarily high levels. While the results will not precisely match those obtained using N-body trees they are sufficiently close to allow us to explore the effects of changing resolution. Figure~\ref{fig:ResStudy} below shows the results of a resolution study carried out in this way on the morphologically segregated luminosity function.

The thickest lines correspond to the case of N-body merger trees drawn from the Millennium Simulation (i.e. used in this paper). The remaining lines were all generated using the exact same model parameters but with merger trees constructed using the \cite{parkinson_generating_2008} algorithm. Of these lines, the thickest has a resolution matched to that of the Millennium Simulation. As can be seen (Fig.~\ref{fig:ResStudy}), it does not produce luminosity functions in precise agreement with those obtained using Millennium Simulation merger trees (since the analytic merger trees are not exactly statistically equivalent to the N-body ones), but they are very close. The thinner lines represent results obtained when the mass resolution in the analytic merger trees is increased (i.e. shifted to smaller masses) by factors of 4 and 16. It can clearly be seen that there is no significant change in the predicted luminosity functions. As such, we conclude that galaxy morphologies are well converged in our model when using trees with the Millennium mass resolution. (This is a consequence of the effects of supernovae feedback, which makes galaxy formation in lower mass halos highly inefficient and, therefore, to have little effect on more massive galaxies.)

\begin{figure}
 \includegraphics[width=80mm,viewport=0mm 55mm 205mm 245mm,clip]{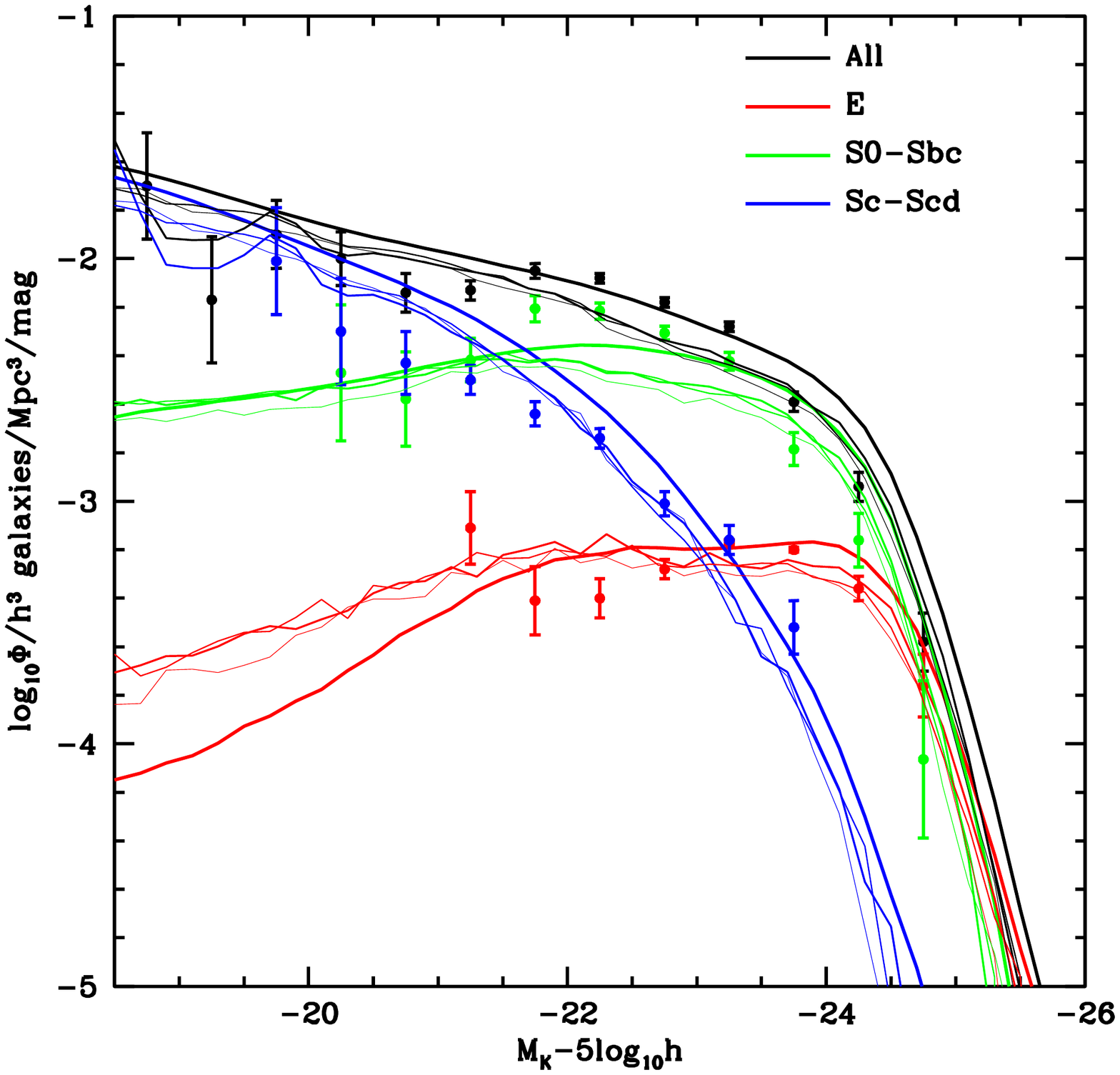}
 \caption{A study of resolution effects on the morphologically segregated luminosity function of Fig.~\ref{fig:BestLF}. The thickest lines show the results obtained utilizing Millennium Simulation merger trees (i.e. as used throughout our work). The remaining three sets of lines were produced using analytically generated merger trees (using the algorithm of \protect\cite{parkinson_generating_2008}). The thickest of these lines correspond to analytic trees with a mass resolution matched to that of the Millennium Simulation, while successively thinner lines have 4 and 16 times better mass resolution.}
 \label{fig:ResStudy}
\end{figure}

\subsection{Alternative Treatment of Disk Instabilties}\label{app:altDiskInstab}

\begin{figure}
 \includegraphics[width=80mm,viewport=0mm 55mm 205mm 245mm,clip]{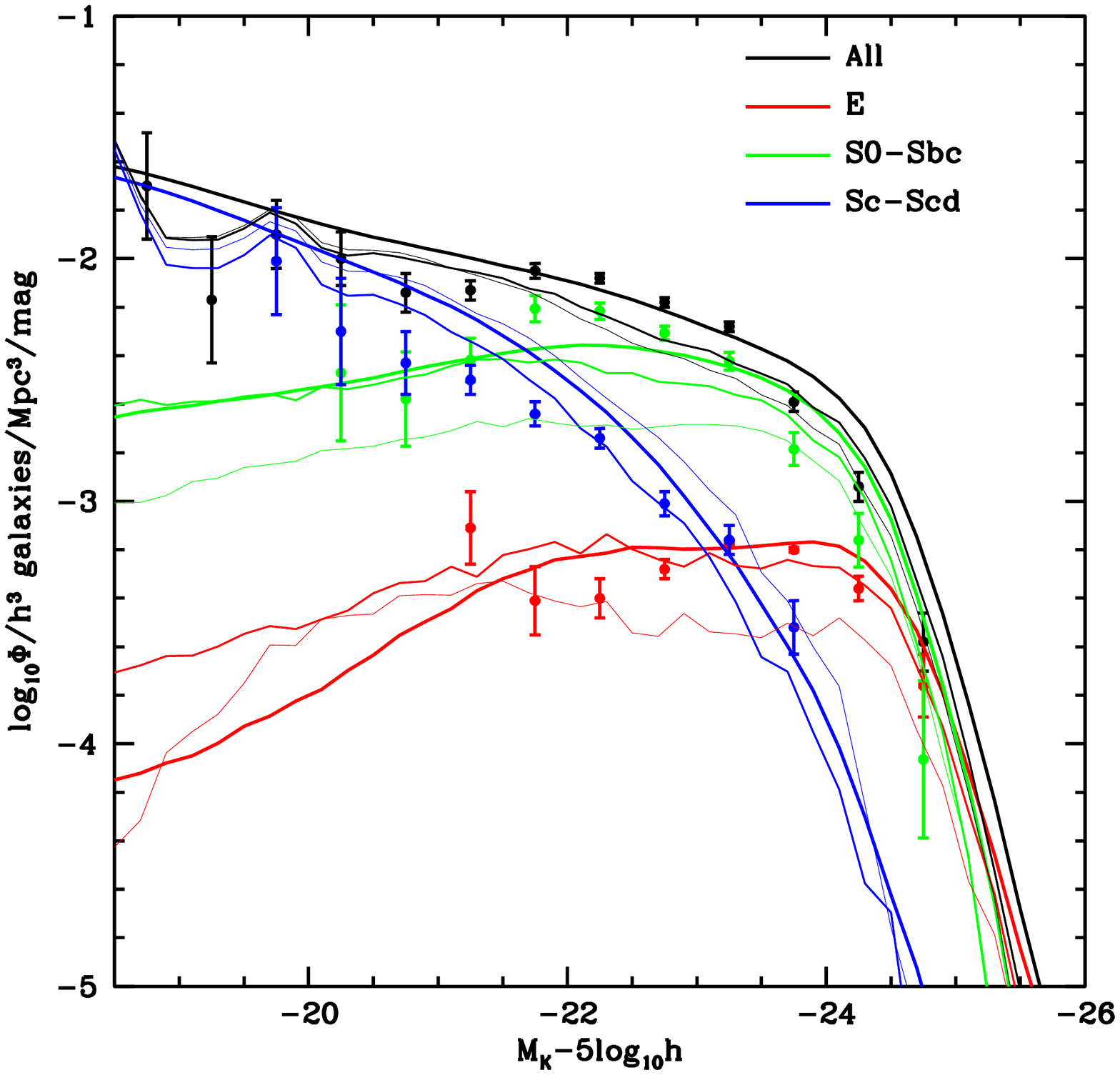}
 \caption{The effects of an alternative disk instability model on the morphologically segregated luminosity function. The thickest lines show results from our standard treatment of disk instabilities applied to merger trees extracted from the Millennium Simulation. Intermediate lines indicate the same standard treatment of instabilities but applied to analytically derived merger trees (using the algorithm of \protect\cite{parkinson_generating_2008}). Finally, the thinnest lines show the results of adopting an alternative treatment of disk instabilities (in which only just enough mass is transferred from an unstable disk to cause it to restabilize) applied to analytically derived merger trees.}
 \label{fig:altDiskInstab}
\end{figure}

The instability of galaxy disks is a difficult process to treat analytically. Our standard approach assumes that any unstable disk will quickly develop a strong bar and destroy itself entirely, turning the galaxy into a pure spheroid. Recent, high resolution hydrodynamical simulations \pcite{agertz_disc_2009}  show that unstable disks experience local instabilities leading to fragmentation and the incomplete destruction of the disk. To at least partially assess the importance of the assumptions we make regarding disk instabilities we have implemented an alternative disk instability treatment in our model in which only just enough mass is transferred from the disk to the spheroid component to restabilize the disk---an approach which has been adopted by other semi-analytic models \pcite{hatton_galics-_2003,croton_many_2006}. Figure~\ref{fig:altDiskInstab} shows the effect of switching to this alternative treatment on the morphologically segregated luminosity function. The thickest lines show results utilizing Millennium Simulation trees with our standard treatment of disk instabilities. Medium thickness lines show the same treatment of instabilities but with analytic merger trees with resolution matched to that of the Millennium Simulation. Finally, the thinnest lines show results of using the alternative treatment of instability events with analytic merger trees.

The net result of switching to the alternative treatment of instabilities is to slightly increase the number of bulgeless galaxies at all luminosities, with a corresponding decrease in the numbers of intermediate and pure spheroid galaxies. 
The changes do not alter the qualitative trends of morphological mix with luminosity. 
However, in order to reproduce quantitatively the observed morphological mix of galaxies at  $M_{\rm K}-5\log h = -23$ would require adjusting
the range of B/T ratios used to segregate galaxies into the three broad morphological classes
(Table~\ref{tb:BTMap}). The sense of the required adjustment for the E (``pure spheroid'') class would be to smaller B/T ratios, which is inconsistent with their observed morphology and may
be a sign that our standard treatment of disk instabilities is more realistic.

\end{document}